\documentclass{aa}  
\usepackage{graphicx}
\usepackage{txfonts}
\usepackage{hyperref}
\hypersetup{
    colorlinks=true,
    linkcolor=blue,
    citecolor=blue,   
    urlcolor=cyan,
    }
\usepackage{orcidlink}

\defcitealias{norazChromosphereQuietSun2026}{Paper~I}

\begin{document}

   \title{Chromosphere of the quiet Sun}
   \subtitle{II. Atmospheric response to small-scale magnetic flux emergence}
   \author{Q. Noraz\inst{1,2}\thanks{E-mail: quentin.noraz@astro.uio.no}, M. Carlsson\inst{1,3} \and G. Aulanier\inst{1,4}
          }

   \institute{Rosseland Centre for Solar Physics, University of Oslo, P.O. Box 1029 Blindern, Oslo, NO-0315, Norway\\
             \and Centre for Mathematical Plasma-Astrophysics, Department            of Mathematics, KU Leuven, Celestijnenlaan 200B, 3001               Leuven, Belgium\\
             \and
             Institute of Theoretical Astrophysics, University of Oslo, P.O.Box 1029 Blindern, Oslo, NO-0315, Norway.\\
             \and
             Sorbonne Université, Observatoire de Paris – PSL, École Polytechnique, Institut Polytechnique de Paris, CNRS, Laboratoire de Physique des Plasmas (LPP), 4 Place Jussieu, 75005 Paris, France\\
             }

   \date{Received 25 March, 2026; Accepted 2 May, 2026}
 
  \abstract
   {The couplings between the photosphere, chromosphere, and corona in the quiet Sun (QS) are governed by a complex interplay between magnetic structuring, heating, mass-loading, and radiative cooling. The current constraints on how this balance responds to variations in small-scale magnetic flux are limited.}
   {We investigate how chromospheric heating and the thermodynamic response of higher atmospheric layers vary as a function of small-scale magnetic flux emergence under QS conditions.}
   {We performed a parametric set of 3D radiative-MHD simulations with the \textit{Bifrost} code, starting from a weakly magnetised quiet-Sun reference model and injecting horizontal magnetic flux of increasing amplitude into the sub-surface convection zone. We analysed the resulting chromospheric dynamics, heating, mass-loading, and coronal response in quasi-static regimes.}
   {Chromospheric temperatures and mechanical heating rise monotonically with increasing magnetic-field strength. Although the fractional contribution of shocks decreases from 23 to 5\%, reconnecting current sheets (CSs) continue to remain steady at about $50\%$. In contrast, the temperature at the base of the corona exhibits a non-monotonic response, reaching a maximum at intermediate magnetic amplitudes and decreasing for the strongest-field case. We show that stronger magnetic-field strength increases chromospheric heating, thereby increasing the coronal-base density through efficient mass-loading, and amplifies radiative losses. These density-driven radiative losses dominate the coronal energy balance and, thus, lead to reduced coronal-base temperatures despite increased heating.}
   {Our results demonstrate the sensitivity of chromospheric structure and dynamics to small-scale flux emergence and its key role in regulating coronal thermodynamics. In particular, this study has revealed a non-monotonic thermodynamic response in the upper atmosphere: stronger heating in the chromosphere can paradoxically lead to lower coronal temperatures as increased mass-loading enhances radiative losses. This result illustrates the chromosphere’s role as a thermodynamic gatekeeper, warranting further investigations of realistic flux-emergence models, as well as surface-to-corona parametrisation across various magnetic configurations, relevant to global solar wind models and space weather forecasts.}


   \maketitle

\section{Introduction}\label{sec:sect1}

The quiet Sun (QS) comprises most of the solar surface and exhibits a variety of small-scale magnetic structures. These are typically classified as network (a supergranular-spaced reticulum of strong kilogauss fields) or internetwork (i.e. a dense collection of weaker flux concentrations in between, see \citealt{bellotrubioQuietSunMagnetic2019} for a review). This magnetic field is sustained via ubiquitous and ephemeral flux emergence, resulting from the solar dynamo \citep{schrijverSustainingQuietPhotospheric1997,martinezgonzalezEMERGENCESMALLSCALEMAGNETIC2009,diaz-castilloEmergenceMagneticFlux2025}.

Magnetic fields at the solar surface are commonly inferred via the Zeeman and Hanle effects. Recent advances have improved the reliability of photospheric line-of-sight magnetic field measurements at the disk centre with high-spatial-resolution instruments \citep[e.g.][]{sinjanMagnetogramsUnderestimateEven2024, nobrega-siverioSmallscaleMagneticFlux2024}. However, retrieving other components of the magnetic field, especially in higher atmospheric layers, remains a significant challenge. In particular, the inversion of magnetic fields in the chromosphere is still highly uncertain due to line-formation complexities and limited diagnostics \citep[e.g.][]{delacruzrodriguezRadiativeDiagnosticsSolar2017}.

In numerical models, the magnetic field amplitude and topology are generally imposed as free parameters \citep{carlssonNewViewSolar2019}. This is critical, as the magnetic field configuration influences the chromospheric dynamics \citep[e.g.][]{nindosDynamicChromosphereMillimeter2022}, as well as its coupling with other layers of the atmosphere and, therefore, the way in which mass and energy are transferred to the solar environment. While several studies have started to explore the impact of various magnetic topologies on the chromosphere \citep{carlssonPubliclyAvailableSimulation2016,martinez-sykoraOriginMagneticEnergy2019,przybylskiChromosphericExtensionMURaM2022,martinez-sykoraChromosphericHeatingLocal2023,przybylskiStructureDynamicsInternetwork2025} as well as active-Sun amplitude flux-emergence \citep{archontisCLUSTERSSMALLERUPTIVE2014,ortizEMERGENCEGRANULARSIZEDMAGNETIC2014,hansteenBombsFlaresSurface2017,hansteenEllermanBombsUV2019}, parametric investigations are still pending for QS conditions.

It has been recurrently shown that the coupling between the solar chromosphere and corona is governed by a delicate balance between heating, mass-loading, and cooling \citep{gudiksenInitioApproachSolar2005a,rempelEXTENSIONMURAMRADIATIVE2017,carlssonNewViewSolar2019}. The coronal temperature alone provides an incomplete measure of the energy budget, and atmospheric coupling must therefore be understood as a joint mass–energy problem. This type of a non-linear feedback loop depends on magnetic geometry and therefore needs to be explored across the diverse configurations present on the Sun.

Magnetic fields are known to modulate chromospheric heating. Observations suggest that acoustic shocks alone might suffice to balance radiative losses in the lower chromosphere of some internetwork regions. However, their contribution likely drops to 30–50\% near plage boundaries \citep{abbasvandObservationalStudyChromospheric2020}, showing that the magnetic field likely regulates the relative weight of heating mechanisms, which remains to be quantified \citep{carlssonNewViewSolar2019}. In \citet[][hereafter \citetalias{norazChromosphereQuietSun2026}]{norazChromosphereQuietSun2026}, we quantified the respective contributions of shock waves and current sheets (CSs) in a Bifrost QS simulation and found that they jointly provide the strongest contribution of the chromospheric heating for a weakly magnetised quiet-Sun simulation (see also \citealt{udnaesCharacteristicsAcousticwaveHeating2025,cherryDecomposingWaveActivity2025} for complementary analysis on similar models). However, understanding how this chromospheric small-scale dynamics and subsequent energy deposition is evolving as a function of the magnetic-field topology and amplitude is still pending.

There is now a broad consensus that heating in the lower solar atmosphere is an intermittent, multi-physics process that can hardly be captured without comprehensive 3D radiative-MHD models (see e.g. \citealt{carlssonNewViewSolar2019,przybylskiStructureDynamicsInternetwork2025,lamarreAvalanchesMagnetohydrodynamicalSimulations2025}; \citetalias{norazChromosphereQuietSun2026}). Such models can indeed provide essential constraints for parameterising this dynamics and subsequent mass-energy transfers in global approaches that employ crude chromospheric prescriptions. In particular, while modelling the chromosphere comprehensively will likely remain out of Space-Weather operational reach for the near future (see e.g. \citealt{brchnelovaCOCONUTMFTwofluidIonneutral2023}), improving the parameterisation of the low solar atmospheric coupling in different magnetic environments is both feasible and increasingly necessary \citep[e.g.][]{vanderholstALFVENWAVESOLAR2014,parentiValidationWaveHeated2022,brchnelovaConstrainingInnerBoundaries2025,wangMHDModellingOpen2026}. 

Building on the work we presented in \citetalias{norazChromosphereQuietSun2026}, we have extended it to a controlled parametric exploration to assess how the chromospheric thermodynamics varies under different QS conditions. Specifically, we aim to mimic idealised small-scale flux emergence with different amplitudes, and explore how the flux injected influences (i) chromospheric heating processes and (ii) the subsequent response of higher layers due to changes in mass and energy transfer.

In Sect.~\ref{sec:sect2}, we present the simulation setup. The behaviour of the flux we inject is described, along with the global impact on the magnetic and temperature structure. We then focus our analysis on changes in the chromospheric heating and dynamics in Sect.~\ref{sec:sect3}, before focussing on the coupling to higher layers in Sect.~\ref{sec:sect4}, where we investigate the changes in temperature and mass-loading. Finally, we discuss the different caveats of our models in Sect.~\ref{sec:sect5} before presenting our conclusions in Sect.~\ref{sec:sect6} and opening up the potential impact and perspective offered by these results.

\section{Parametric setup of the experiment}\label{sec:sect2}

\subsection{Construction of the models}\label{sec:setup}

We used the \textit{ch012023} simulation (hereafter \textit{Ref}) presented in \citetalias{norazChromosphereQuietSun2026} as the reference run of our controlled parametric exploration. We solved the 3D time-dependent, resistive MHD equations with the \textit{Bifrost} code \citep{gudiksenStellarAtmosphereSimulation2011}, which models the solar atmosphere from the sub-surface convection zone (CZ) to the low corona, including the photosphere, chromosphere, and transition region (TR), in a Cartesian box. This reference run reproduces QS conditions with an open magnetic topology characteristic of a coronal hole and self-consistently sustained by a local dynamo in the CZ (i.e. no magnetic flux is injected through the boundaries).

To mimic different idealised and small-scale flux-emergence configurations in the QS, we duplicated this reference run into two additional simulations, \textit{ch012023\_by200} and \textit{ch012023\_by800}, in which a uniform horizontal magnetic-flux sheet ($B_y = 200$ and $800$~G, respectively) is injected into convective upflows through the bottom boundary. These values were chosen to ensure efficient emergence through the convection zone, while spanning average photospheric fields, characteristic from weak to strong flux emergence under quiet-Sun conditions (see Sect.~\ref{sec:relax}. All three runs share the same numerical setup, which we briefly summarise here (see \citetalias{norazChromosphereQuietSun2026} for more details).

The simulations were computed on a $512^3$ grid spanning 12~Mm in both horizontal directions, with periodic boundary conditions and a constant horizontal resolution of 23~km (prefix \textit{ch012023}). Vertically, the domain extends from 2.5~Mm below to 8~Mm above the mean solar surface ($\tau_{500} = 1$) with non-uniform spacing: 30~km at the base of the CZ, 14–12~km in the photosphere and chromosphere, then up to 70.5~km at the coronal top. The lower boundary allows for inflows with entropy adjusted to maintain an effective temperature close to $\sim 5780$~K, while the upper boundary applies an open characteristic-boundary scheme (\citealt{gudiksenStellarAtmosphereSimulation2011}, see also \citealt{tarrSimulatingPhotosphericCoronal2024}). In \textit{ch012023\_by200} and \textit{ch012023\_by800} (hereafter referred to as \textit{By200} and \textit{By800}, respectively), the magnetic-flux sheet injection begins at $t = 140$~min, and continues for the remainder of the simulated time.

\subsection{Behavior of the flux emergence}\label{sec:FEbehav}

\begin{figure*}
    \begin{center}
        \includegraphics[width=\linewidth]{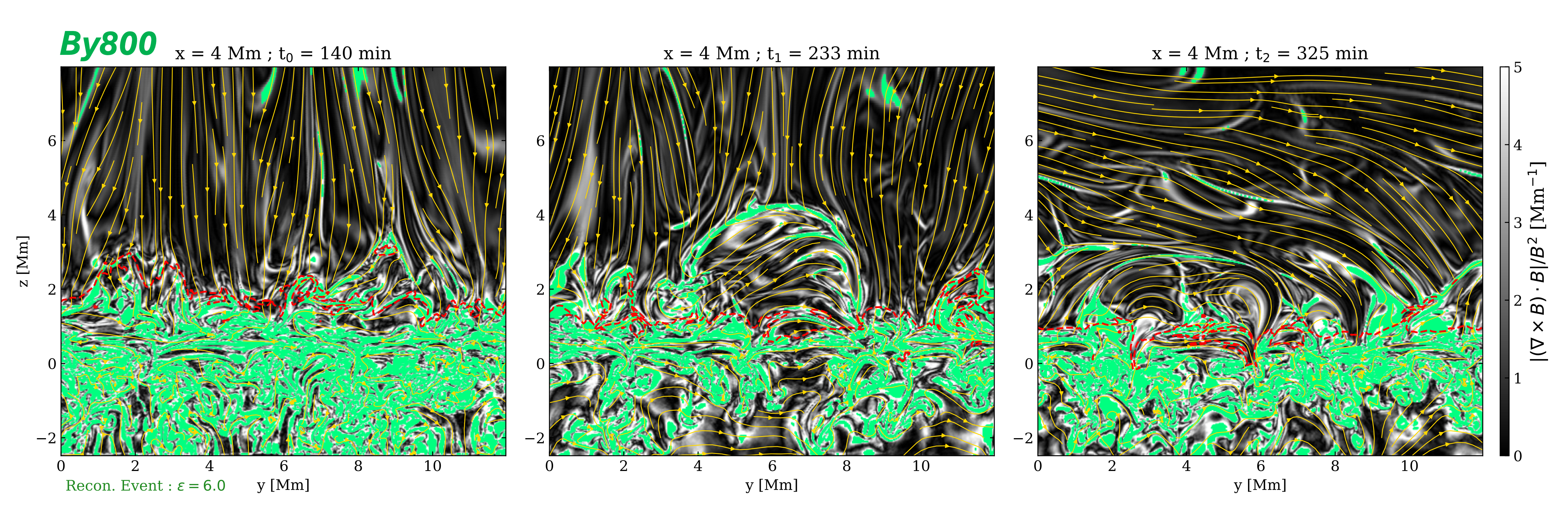}
    \end{center}
    \caption{Magnetic-field evolution during the three main phases of the \textit{By800} experiment for $t_0=140$~min, $t_1=233$~min and $t_2=325$~min. The panels show the normalised parallel current, $|\nabla\times\mathbf{B}\cdot\mathbf{B}|/|\mathbf{B}|^2$ (grayscale), and the reconnecting current sheets (CSs; green), identified using the criterion of Eq.~\ref{eq:alpha_crit}. Magnetic-field lines are shown with yellow streamlines, and the $\beta=1$ surface is drawn with a dashed red line. We note how the height of the latter has moved from the initial $t_0$ to the quasi-static timestep $t_2$, due to the injection of emerging magnetic-loop structures, increasing the volume filled by a subsequently reconnecting CS. The associated movie is available online.}
    \label{fig:FEbehav}
\end{figure*}

\begin{figure}
    \begin{center}
        \includegraphics[width=\linewidth]{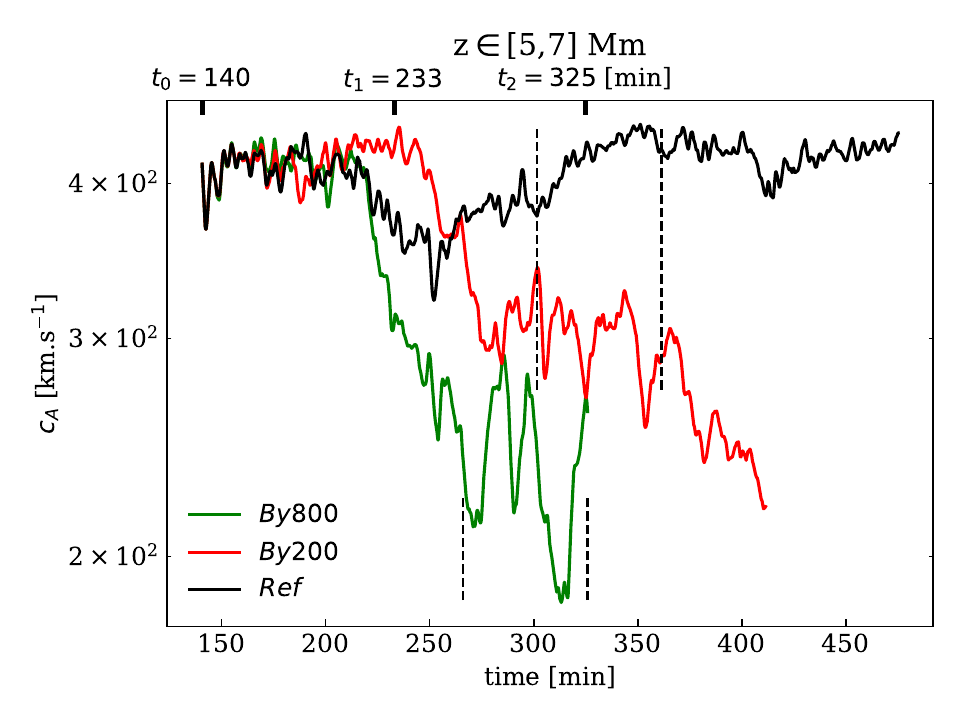}
    \end{center}
    \caption{Temporal evolution of the Alfvèn speed $c_A= \sqrt{B^2/4\pi\rho}$, spatially averaged from 5 to 7 Mm above the photosphere, for \textit{Ref} (black), \textit{By200} (red), \textit{By800} (green). The vertical dashed lines mark the time intervals used for the analyses presented in the next sections.}
    \label{fig:atmoCA}
\end{figure}

Magnetic flux is continuously injected from the bottom boundary, leading to a temporal evolution of the magnetic topology in the atmosphere above. To illustrate the morphological evolution, Fig.~\ref{fig:FEbehav} shows the \textit{By800} model at three representative timesteps $t_0$, $t_1$, and $t_2$ (see also Fig.~\ref{fig:atmoCA}). The normalised parallel current,
\begin{equation}
    \frac{|\nabla \times \mathbf{B} \cdot \mathbf{B}|}{|\mathbf{B}|^2} > \frac{1}{\epsilon\, ds},
    \label{eq:alpha_crit}
\end{equation}
with $\epsilon=6$ and $ds=\max(dx,dy,dz)$, is used to identify reconnecting CSs following the method of \citetalias{norazChromosphereQuietSun2026}.

In the initial state ($t_0$ ; left panel), the field is predominantly vertical (yellow lines), consistently with the imposed average vertical flux of 2.5~G, representative of quiet-Sun coronal-hole conditions \citep{harveyMagneticMeasurementsCoronal1982, zwaanElementsPatternsSolar1987}. This experiment is initialised from the \textit{Ref} model, whose detailed dynamics are discussed extensively in \citetalias{norazChromosphereQuietSun2026}. Most CSs (green patches) are concentrated below the $\beta=1$ surface (dashed red line), horizontally aligned in the photosphere and following the convective flows dynamics in the CZ.

During what we will refer to as the 'transient phase', horizontal magnetic flux is advected upwards from the lower boundary and begins to emerge into the chromosphere. At $t_1$, in the middle panel, a dome-shaped structure between $4<y<8$~Mm forms as the field rises, bounded by reconnecting CSs (green) that develop at the interface, due to interaction of the emerging field with the ambient one (see the corresponding animation, and also e.g. \citealt{archontisCLUSTERSSMALLERUPTIVE2014, ortizEMERGENCEGRANULARSIZEDMAGNETIC2014}). In the CZ, we note large areas (i.e. of a few~Mm) where CSs are now absent and the normalised current is weak (darker regions). These correspond to upflows, where the uniform and untwisted horizontal flux is injected, as can be seen in the animation. Furthermore, the amplitude of the magnetic field injected at $z=-2.5$~Mm ($By=800$~G in this case) is strong enough so that the CS formation is notably limited up to the photosphere, in comparison to the initial state, due to the subsequent increase in the Lorentz force. This leads to a preferential location of CSs in downflow lanes, where the kinetic energy density is stronger.

Once the emerging flux reaches the top of the domain and starts to leave through the open boundary conditions, a quasi-equilibrium settles as magnetic flux now both enters and leaves the domain, from the bottom and top boundaries, respectively. During this 'quasi-static' phase ($t_2$ ; right panel), the injected flux has now spread in the whole domain, notably producing a substantial horizontal magnetic component, extending in the chromosphere and up to the top of the domain (see Fig.~\ref{fig:FEbehav}). The $\beta=1$ surface has moved downwards as the magnetic pressure increased, while the volume occupied by CSs has expanded. This behaviour is further quantified in Sect.~\ref{sec:ShCs}.

\begin{figure}
    \begin{center}
        \includegraphics[width=0.9\linewidth]{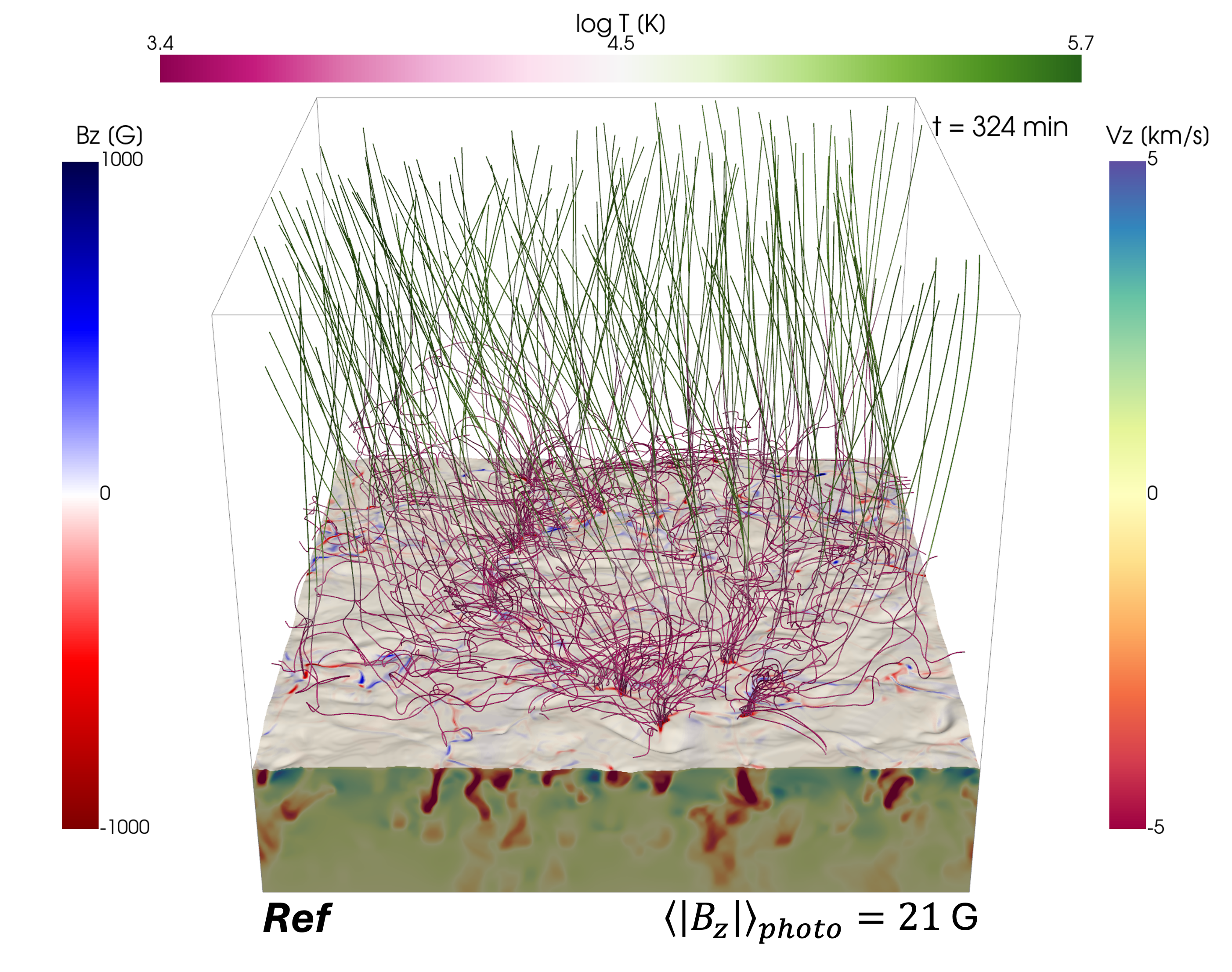}
        \includegraphics[width=0.9\linewidth]{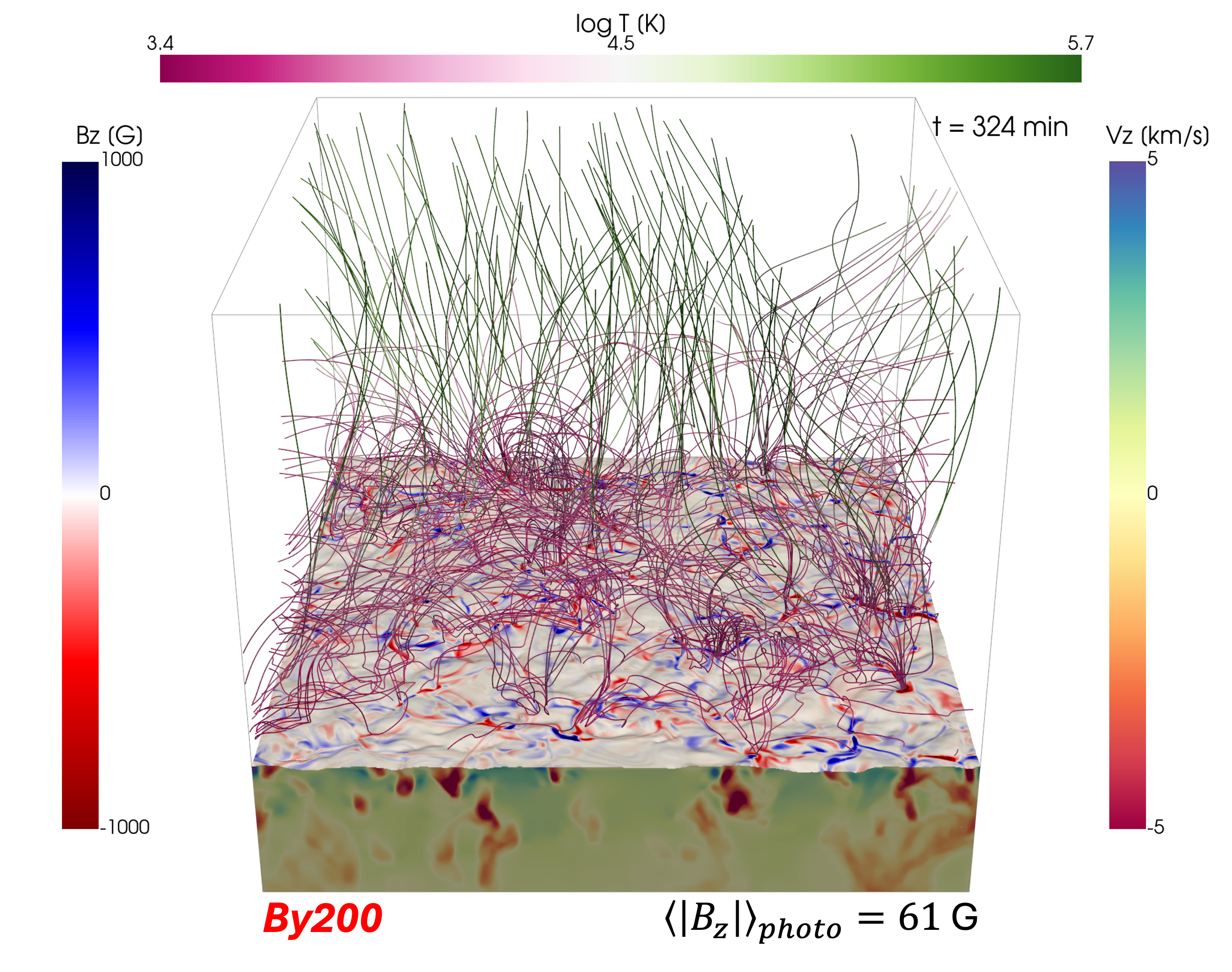}
        \includegraphics[width=0.9\linewidth]{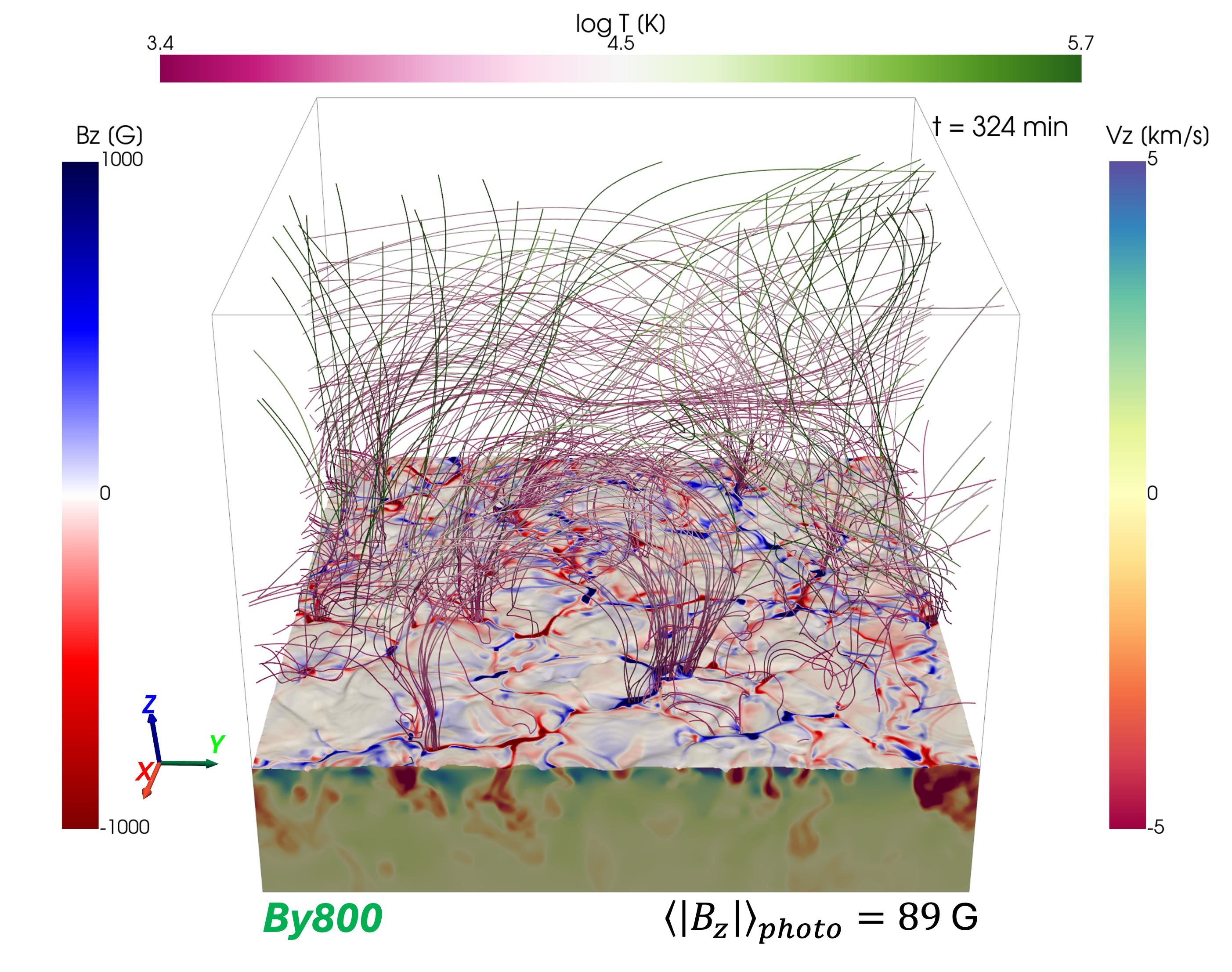}
    \end{center}
    \caption{3D visualisations of magnetic and thermodynamic structures in the three simulations: \textit{Ref} (top), \textit{By200} (middle), and \textit{By800} (bottom), shown during their quasi-static phases at t = 324~min. The corrugated horizontal surface marks the $\tau_{500}=1$ layer, coloured by the vertical magnetic field $B_z$. The vertical side panels show the convective velocity $v_z$ in the upper convection zone, while magnetic field lines, seeded from a uniform $15\times15$ grid at $z=3$~Mm, are coloured by temperature. This transitions from pink in the chromospheric temperature minimum ($\sim$4000~K), to white in the TR ($\sim$40,000~K), and green in the low corona ($\sim$500,000~K). The panels illustrate the progressive emergence of horizontal flux, from the nearly vertical, coronal-hole-like topology of \textit{Ref}, through loop-dominated \textit{By200}, to the strongly inclined, horizontally dominated configuration of \textit{By800}. Associated movies are available online.}
    \label{fig:3Drend}
\end{figure}

To further quantify the temporal evolution, we compute and illustrate in Fig.~\ref{fig:atmoCA} the spatial average of the alfvèn speed $c_A=\sqrt{B^2/4\pi\rho}$ from 5 to 7 Mm above the photosphere. First, a pronounced decrease in $c_A$ is observed in both \textit{By200} (red) and \textit{By800} (green) starting at $t\sim 245$ and 225~min, respectively. This marks the onset of magnetic flux emergence into the averaged volume. The decrease in $c_A$ further indicates that the density, $\rho$, increases faster than $B^2$ on average. This shows that denser material is loaded up as the magnetic structures we inject rise into the atmosphere, which will be further analysed in Sect.~\ref{sec:sect4}. The earlier Alfvén speed decrease in the \textit{By800} case is then consistent with its stronger imposed field, and the corresponding increase in magnetic buoyancy, $F_b$. To first order, $F_b$ scales with the field amplitude as $F_b = g\delta\rho/\rho \propto \beta^{-1}$, where $g$, $\rho$, and $\beta$ are the gravitational acceleration, density, and plasma-$\beta$ parameter, respectively \citep{moreno-insertisNonlinearTimeevolutionKinkunstable1986, cheungFluxEmergenceTheory2014}.

Once the magnetic flux begins to both enter and leave the mean volume continuously, the injected field has reached the upper part of the domain and the system enters the quasi-static phase in which the decrease in the Alfvén speed starts to saturate. This occurs from $t\sim 300$ and 260~min in \textit{By200} and \textit{By800}, respectively. The vertical dashed lines in Fig.~\ref{fig:atmoCA} mark the one-hour segments within this phase that are used for the time-averaged analysis presented in the next sections.

\subsection{Relaxed structure}\label{sec:relax}

We present in Fig.~\ref{fig:3Drend} a 3D rendering of the three runs at $t_2=325$~min, namely, three solar hours after we started the flux-emergence experiment in both \textit{By200}, and \textit{By800}. This allowed us to compare their respective magnetic and thermal configurations during the quasi-static phase (time ranges between dotted lines of Fig.~\ref{fig:FEbehav}). 

We first note the \textit{Ref} run exhibits a predominantly vertical topology consistent with quiet coronal-hole conditions analysed in \citetalias{norazChromosphereQuietSun2026} and used as initial condition for \textit{By200} and \textit{By800}. In contrast, the \textit{By200} and \textit{By800} cases display progressively more inclined ambient fields and an extended magnetic-loops network, reaching up to 5~Mm. 

We select a solar-hour period over which we will perform temporally averaged analysis (see Fig.~\ref{fig:atmoCA}). We select this period from $t=301$ to 361~min for both \textit{Ref} and \textit{By200} runs, while we select this period from $t=266$ to $325$~min in \textit{By800} (see the dashed lines in Fig.~\ref{fig:atmoCA}). The corresponding mean unsigned photospheric field strengths are $\langle |B_z| \rangle = 21$, 61, and 89~G for the \textit{Ref}, \textit{By200}, and \textit{By800} cases, respectively, once averaged over the solar hour over the $\tau_{500nm}=1$ surface. The three runs presented in this paper hence span different QS typical configurations, from weakly magnetised quiet-Sun conditions to strong small-scale emergence episodes \citep{bellotrubioQuietSunMagnetic2019, gosicBifrostModelsQuiet2025}.

Besides the expected change in topology, we also note the temperature structure changes between the cases as well. While \textit{Ref} and \textit{By200} show coronal temperatures exceeding 100~kK (green shades), \textit{By800} regularly exhibits cooler temperatures in these upper layers (pink tones). This indicates a substantial modification of the thermal structure as well.

\begin{figure*}
    \sidecaption
    \includegraphics[width=12cm]{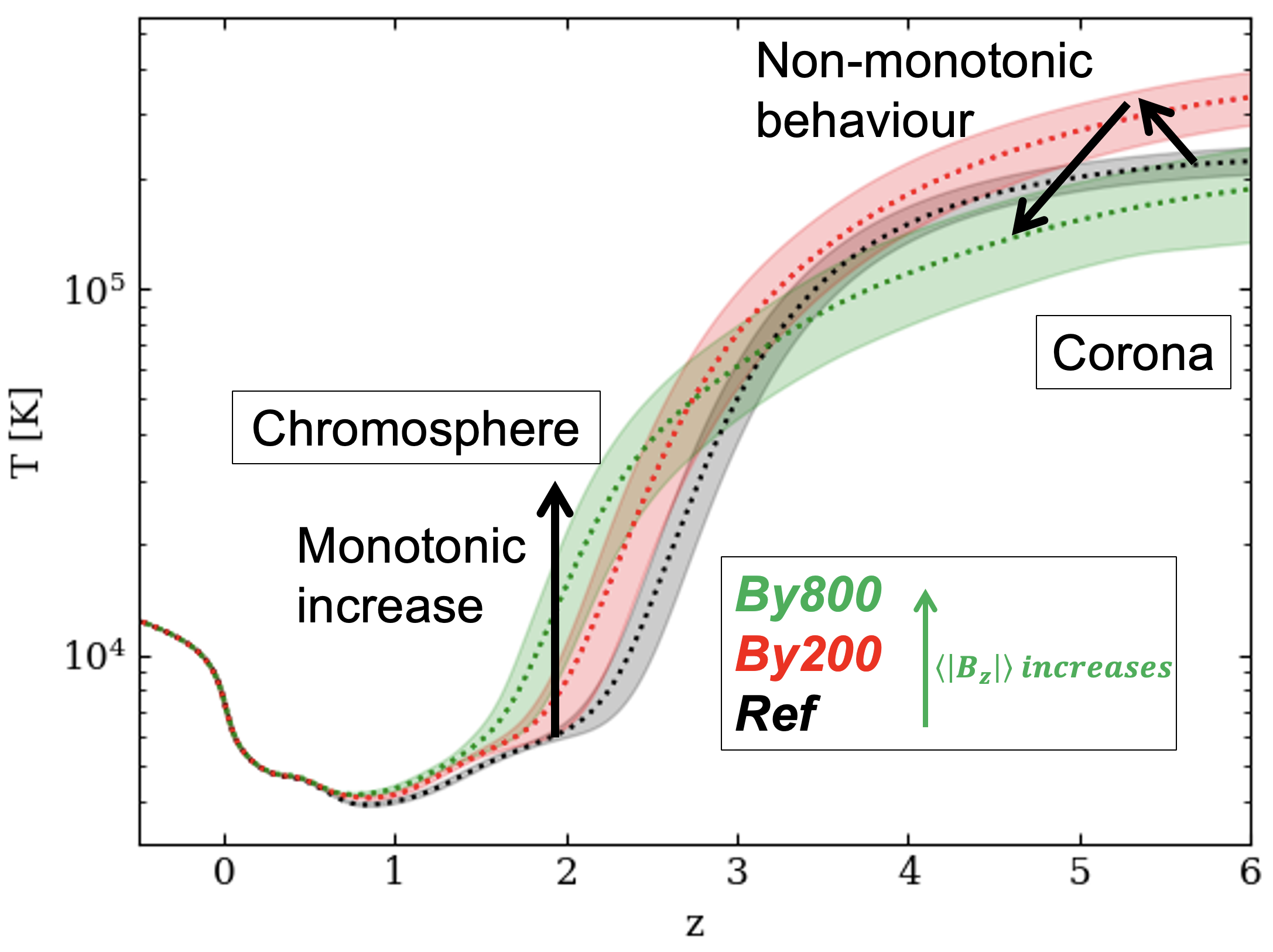}
    \caption{Comparison of the temperature profiles as a function of height, among \textit{Ref}, \textit{By200} and \textit{By800} in black, red and green, respectively. These vertical profiles, and the ones from the following figures, are averaged over one hour of solar time, illustrated in Fig.~\ref{fig:atmoCA}. Horizontal and temporal averages are illustrated with dotted lines, while the envelope indicates $\pm 1$ standard deviation in time. We use black arrows to underline the different noteworthy behaviours here; namely, the chromospheric temperature increase as a function of the magnetic amplitude we inject, along with a non-monotonic response at the base of the corona.}
    \label{fig:tempProfs}
\end{figure*}

To quantify this change, we present the different temperature profiles of our models in Fig.~\ref{fig:tempProfs}. We first note that the intermediate \textit{By200} case (red) is consistently hotter than the \textit{Ref} one (dark) at all heights, from the bottom of the chromosphere and above. However, this is not the case of the strong \textit{By800} run (green), which exhibits the hottest chromosphere, but also the coolest averaged temperature at the base of the corona. In the following, we aim to understand these different behaviours, first focussing on the chromosphere in Sect.~\ref{sec:sect3} and its coupling with higher layers in Sect.~\ref{sec:sect4}.

We also see that the TR, defined here on this plot as the region between both chromospheric and coronal temperature plateaus, is wider as we increase the magnetic flux amplitude we inject. However, we ought to remain careful when horizontally averaging a corrugated surface such as the TR (see also \citetalias{norazChromosphereQuietSun2026}) and we further note that we scarcely end up reaching a so-called flat coronal plateau in both the \textit{By200} (red) and \textit{By800} (green) cases. This visual TR broadening in height acknowledges the change in magnetic topology we see in Fig.~\ref{fig:3Drend}. Indeed, the quasi-uniform vertical topology of \textit{Ref} leads to similar TR heights over the horizontal extent of the box and, hence, a thinner averaged TR, while \textit{By800} exhibits a network of magnetic loops and concentrations, subsequently corrugating the TR over a broader range of heights (see e.g. \citealt{gabrielMagneticModelSolar1976}).

\section{Chromospheric heating}\label{sec:sect3}

\subsection{Impact on the mechanical heating}

\begin{figure}
    \begin{center}
        \includegraphics[width=\linewidth]{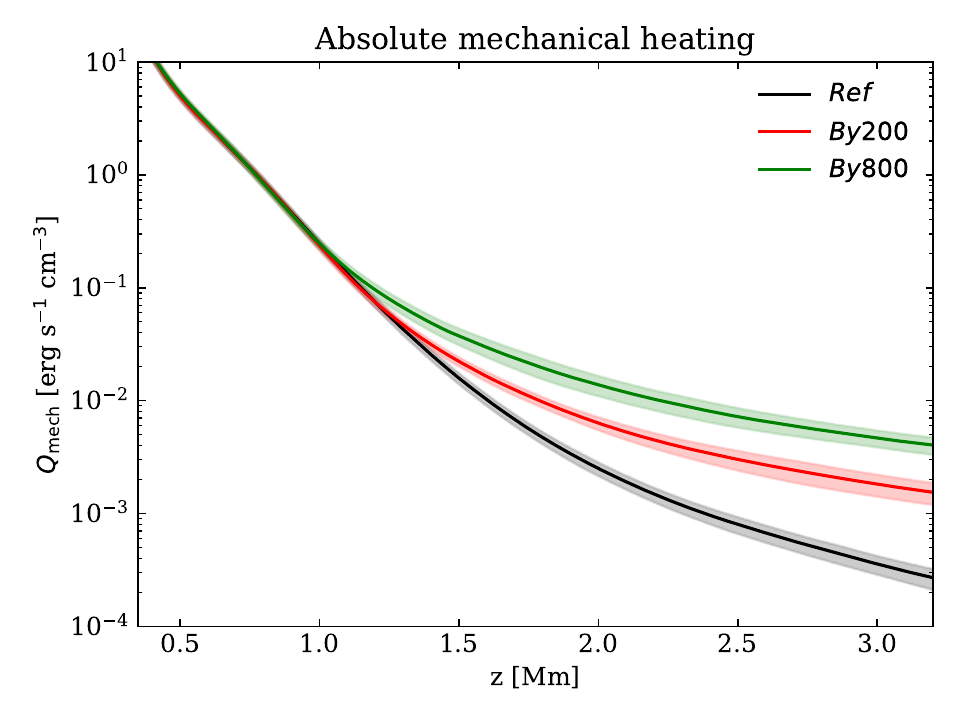}
    \end{center}
    \caption{Comparison of the mechanical heating profiles $Q_{\rm mech}=Q_\nu+Q_\eta+Q_{\rm comp}$ as a function of height, among \textit{Ref}, \textit{By200}, and \textit{By800} in black, red, and green, respectively, averaged horizontally in space and over one solar hour in time. The envelope indicates $\pm 1$ standard deviation in time during the solar-hour average.}\label{fig:Qmech}
\end{figure}
 
Fig.~\ref{fig:Qmech} shows vertical chromospheric profiles of the mechanical heating $Q_{\rm mech}=Q_\nu+Q_\eta+Q_{\rm comp}$, where $Q_{\rm comp}$ is the positive compression contribution from the compressible term $Q_{\rm p\nabla v}=-p\mathbf{\nabla}\cdot\mathbf{v}$, $Q_\nu$ the viscous heating, and $Q_\eta$ the ohmic heating. All simulations exhibit a steep decrease in $Q_{\rm mech}$ with height, due to the steep stratification in this region. 

Yet the runs with larger injected $B_y$ (\textit{By200} and \textit{By800}) show a clear and systematic enhancement of the heating above $z\sim1$~Mm. The difference becomes increasingly pronounced as we go higher in the chromosphere, where the magnetic field is more dominant. This trend suggest that magnetic structuring regulates how mechanical energy is converted into heat and it is consistent with the observational picture where upper-chromospheric temperatures rise with magnetic strength (see e.g. \citealt{withbroeMassEnergyFlow1977,frohlichSolarRadiativeOutput2004,abbasvandChromosphericHeatingAcoustic2020}). Indeed, the profiles point to a more efficient heating as the amplitude of the chromospheric magnetic field increase and we now aim to investigate which processes are at the origin of such behaviour in our set of simulations.

\subsection{Shocks and current sheets individual contributions}\label{sec:ShCs}

\subsubsection{Tracking}

Chromospheric temperatures and dynamics are sustained by a combination of upwardly propagating acoustic waves \citep{biermannUberUrsachenHohen1947,schwarzschildNoiseArisingSolar1948,schmiederWavesLowSolar1979}, magneto-acoustic waves, Alfvén waves \citep{jessMultiwavelengthStudiesMHD2015}, and magnetic reconnection triggered by footpoint shuffling from convective motions \citep{parkerTopologicalDissipationSmallScale1972,parkerMagneticNeutralSheets1983,cargillFineStructureNanoflareheated1993}. These processes have been described in detail in the works of \citet{gudiksenInitioApproachSolar2005,carlssonPubliclyAvailableSimulation2016,carlssonNewViewSolar2019,hansteenNUMERICALSIMULATIONSCORONAL2015,hansteenEllermanBombsUV2019,finleyStirringBaseSolar2022,przybylskiChromosphericExtensionMURaM2022}.

In high-Reynolds and high-Lundquist-number regimes such as the solar chromosphere, these processes are further expected to drive heating at small scales, such as strong velocity $\nabla \cdot \mathbf{v}$ and magnetic field $\nabla \times \mathbf{B}$ gradients, manifesting as shocks and reconnecting CSs, respectively. Following the methodology presented in \citetalias{norazChromosphereQuietSun2026}, we detected shocks and CSs according to the sonic-compression,
\begin{equation}
    -\nabla \cdot \mathbf{v} > \frac{c_s}{\epsilon\, ds},
    \label{eq:cs_crit}
\end{equation}
and normalised-parallel-current criteria presented in Eq.~\ref{eq:alpha_crit}. $c_s$ is the sound speed, $\epsilon = 6$ is tuned to strictly isolate nonlinear dissipation (e.g. shocks) from linear wave propagation (see \citetalias{norazChromosphereQuietSun2026} for further details on the calibration).

\begin{figure*}
    \begin{center}
        \includegraphics[width=0.33\linewidth]{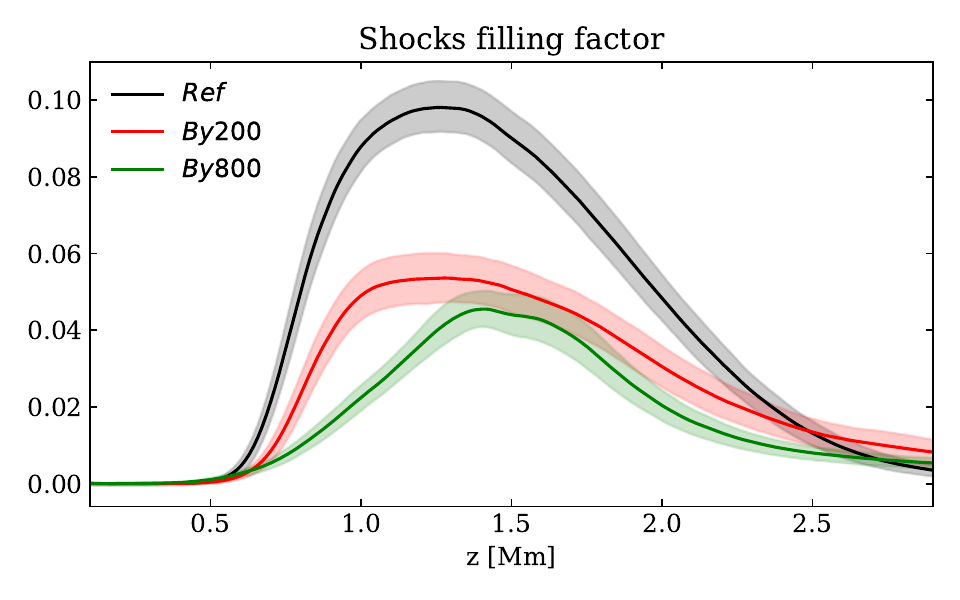}
        \includegraphics[width=0.33\linewidth]{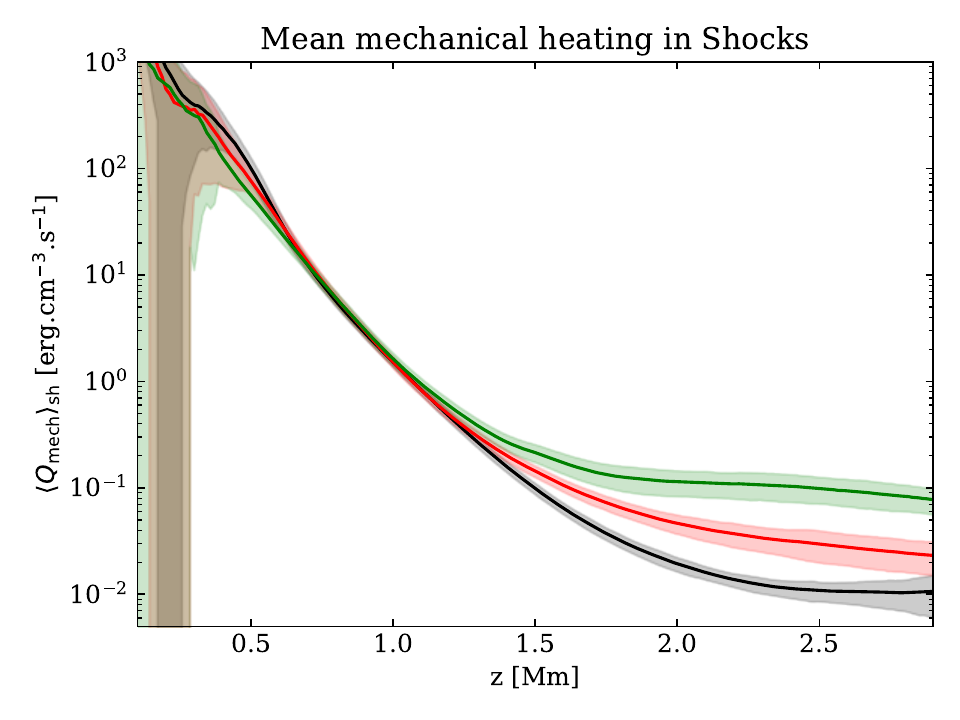}
        \includegraphics[width=0.33\linewidth]{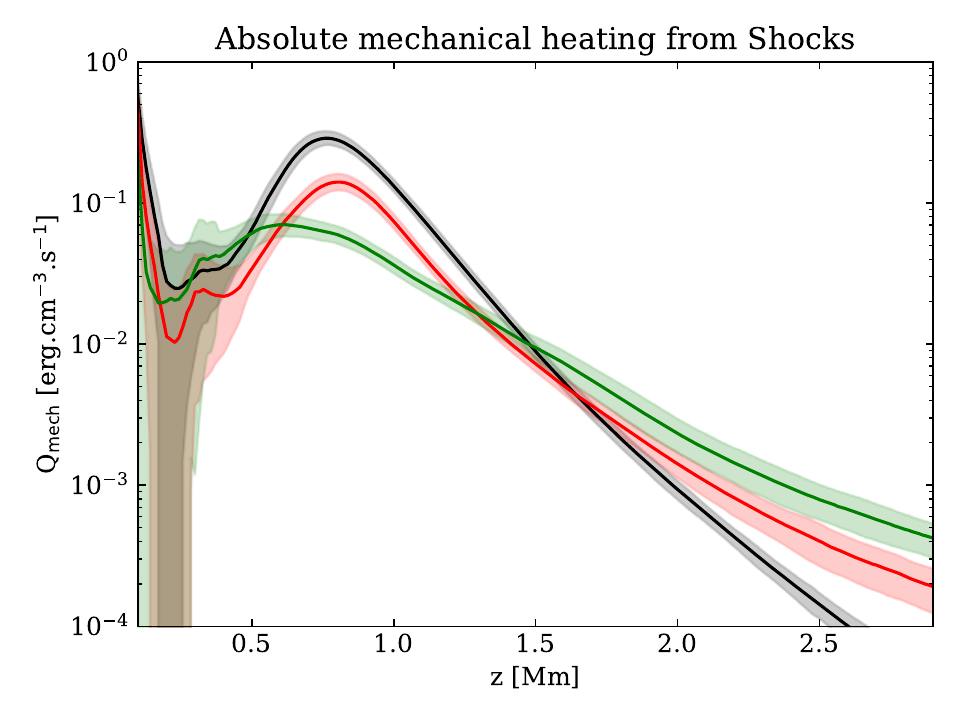}
        \includegraphics[width=0.33\linewidth]{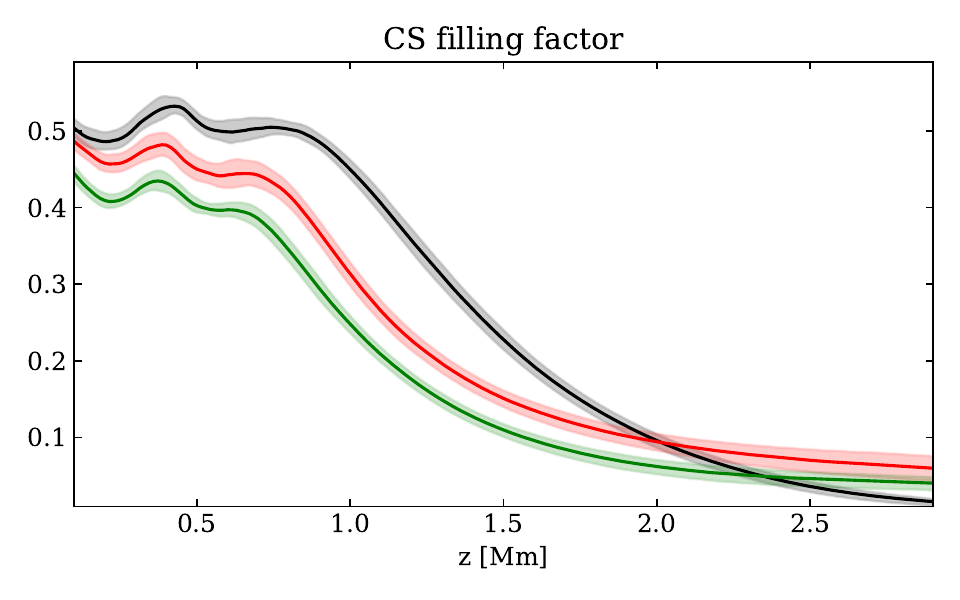}
        \includegraphics[width=0.33\linewidth]{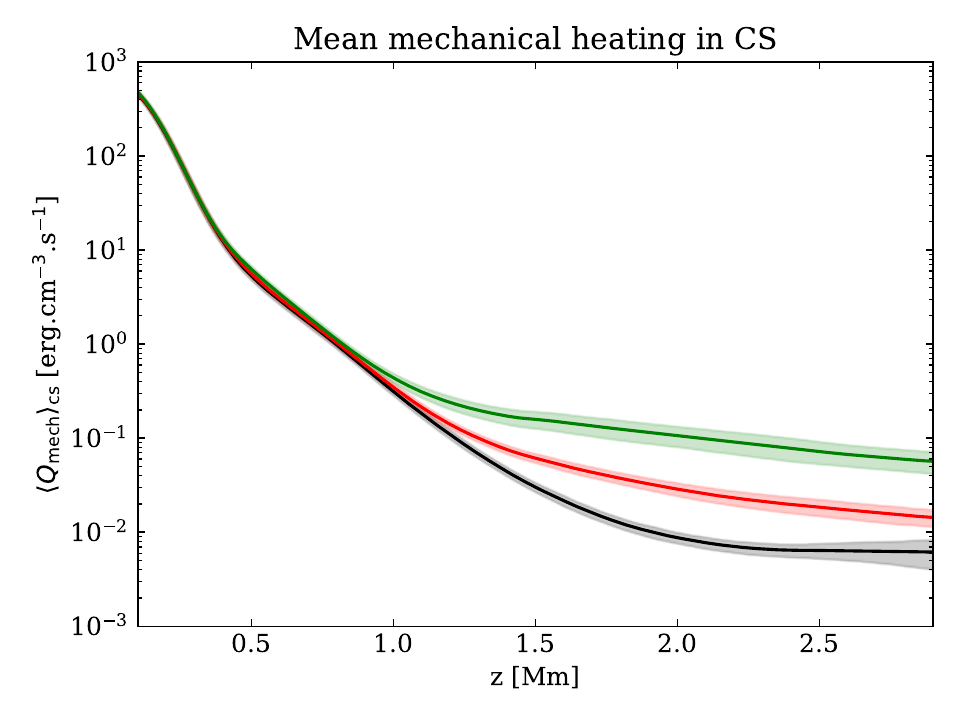}
        \includegraphics[width=0.33\linewidth]{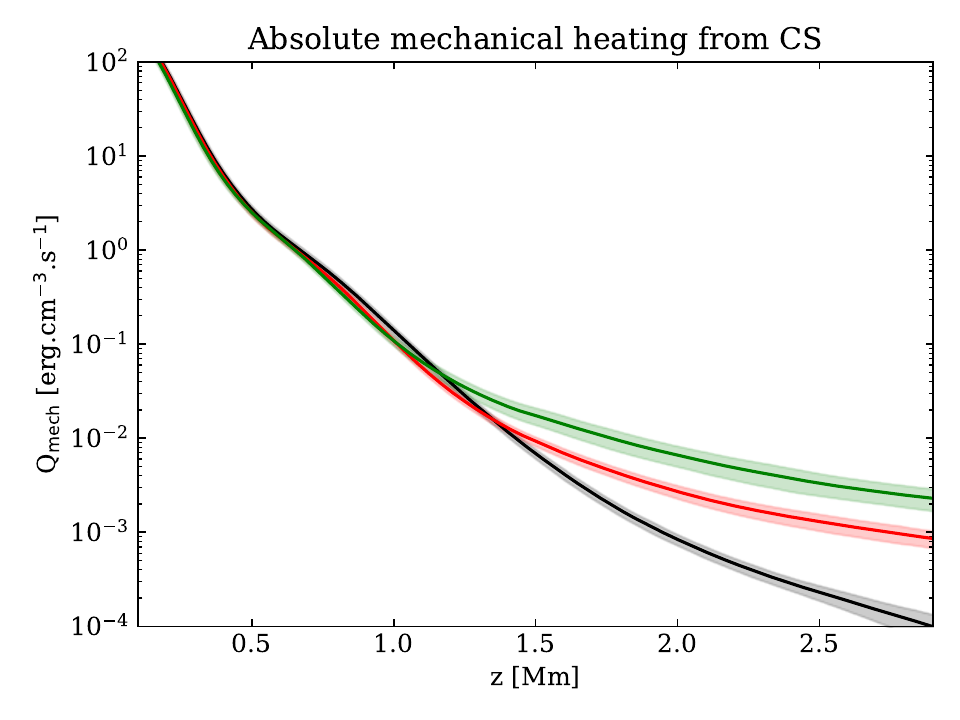}
    \end{center}
    \caption{Shock and CS thermodynamics. Comparison of different profiles as a function of height, among \textit{Ref}, \textit{By200}, and \textit{By800} in black, red, and green, respectively, averaged horizontally in space and over one solar hour in time. The left and middle columns illustrate the filling factors and the mean local mechanical heating at the process location, respectively. The right column panels illustrate the absolute contribution of these processes over the whole horizontal extent, corresponding to the product of the mean local contribution $\langle Q_{\rm mech}\rangle_{x}$ with the corresponding filling factor. Top and bottom rows present it for shocks and CSs, respectively. The envelope indicates $\pm 1$ standard deviation in time.}\label{fig:ShCsHeating}
\end{figure*}

Figure~\ref{fig:ShCsHeating} summarises how shocks, CSs, and their associated mean mechanical heating respond to the imposed horizontal field. The left panels show the vertical evolution of their filling factors, defined as the fractional number of grid cells labelled as shocks. The middle panels display the mean mechanical heating $\langle Q_{\rm mech}\rangle_x$ locally for each process $x=\{sh,cs\}$, namely, averaged over cells labelled as shocks or CSs only, respectively. Finally, the right column illustrates the corresponding absolute contribution over the whole horizontal extent, which corresponds to the product of $\langle Q_{\rm mech}\rangle_x$ with the corresponding filling factor.

\subsubsection{Shocks}

In the top-left panel, all simulations show shock formation starting at $z\simeq 0.5$~Mm, as expected from shock formation in the low solar atmosphere. At this height, all runs still exhibit a high-beta regime, so we do not expect a substantial change in the filling factor as a function of the amplitude of the magnetic field, as the latter is not dynamically dominant yet. We refer to \citetalias{norazChromosphereQuietSun2026} for the dedicated analysis of the evolution with height and set our focus here on its consistent decrease in the chromosphere ($0.5\lesssim z\lesssim 2.5$~Mm) as the injected $B_y$ increases between the different runs. Several mechanisms likely contribute. First, the Lorentz force increasingly counteracts the steepening of upwardly propagating compressive fronts, which reduces shock formation from a dynamical standpoint. Second, the higher chromospheric temperatures in the magnetised runs raise the sound speed $c_s\propto T^{1/2}$, making it more difficult for waves to reach the supersonic regime required for shocks. Third, the magnetic structure of the chromosphere changes (see Fig.~\ref{fig:3Drend}), and more especially the plasma-$\beta$=1 surface is pushed downwards as the magnetic field amplitude increases. This causes the acoustic–magnetic mode conversion to occur earlier during the ascent of wave packets. More wave energy is therefore converted into the fast magnetic mode, which shocks less readily because the Alfvén speed increases. Fourth, we can also mention the deflection of slow-mode acoustic waves by coherent loop structures present in \textit{By200} and \textit{By800}, as the top of which now reaches low-beta regime heights (see also Sect.~\ref{sec:horiz}). This effect will be enhanced as $\beta$ is lower, hence, as the amplitude of the magnetic field is stronger. Determining the relative influence of these processes will require a dedicated analysis, which has been left for future work (see also \citealt{udnaesCharacteristicsAcousticwaveHeating2025,enerhaugIdentifyingMagnetohydrodynamicWave2025,cherryDecomposingWaveActivity2025}).

In the top-middle panel of Fig.~\ref{fig:ShCsHeating}, we examine the evolution of the mean shock-associated heating, $\langle Q_{\rm mech}\rangle_{\rm sh}$. We find that it initially decreases in the lower atmosphere ($z < 0.5$~Mm) as $B_y$ increases. Because this quantity is restricted to shocked cells, this trend does not reflect changes in the filling factor, but instead indicates a local reduction in shock intensity during their formation. Force-balance diagnostics show that the Lorentz force becomes the dominant counteracting term along the propagation direction at these heights, partially inhibiting front steepening in the early stages of shock development.

At greater heights, this trend reverses. Although shocks are less frequent in the \textit{By800} case, those that do form release significantly more mechanical energy once they reach $z \gtrsim 1$~Mm, and $\langle Q_{\rm mech}\rangle_{\rm sh}$ increases with $B_y$. This behaviour becomes even more apparent when considering the absolute contribution of shocks in the top-right panel. The decrease in shock filling factor with increasing $B_y$ leads to a reduction in their total contribution to $Q_{\rm mech}$ in the lower chromosphere ($0.5 \lesssim z \lesssim 1.5$~Mm). 

\begin{figure*}
    \begin{center}
        \includegraphics[width=\linewidth]{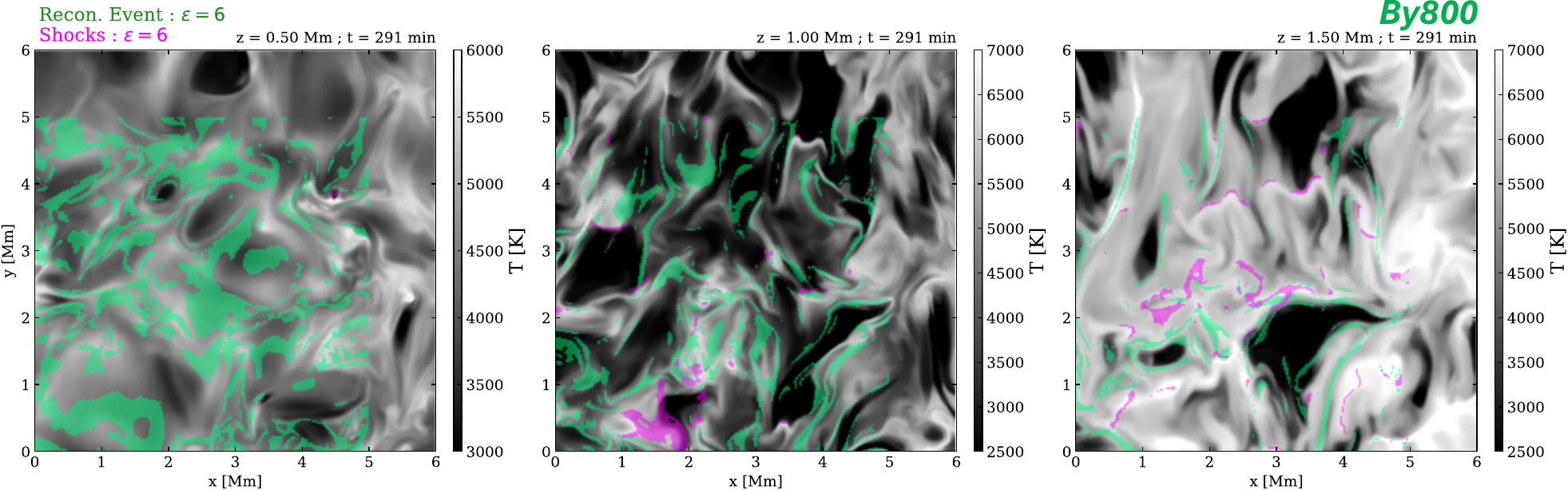}
    \end{center}
    \caption{Shocks (purple) and CSs (green) interplay with temperature structures (greyscale) in the chromosphere of \textit{By800}. A zoom-in on a $6\times 6$~Mm$^2$ area is proposed to focus on small-scale dynamics. Shock and CS overlays are only considered on a $5\times 5$~Mm$^2$ portion, to further illustrate the overlap between them and temperature structures. The $\epsilon$ value specified here refers to the calibration of shocks and CS detections presented in \citetalias{norazChromosphereQuietSun2026}. The associated movie is available online.}\label{fig:HorizLoc}
\end{figure*}

However, in the upper chromosphere ($1.5 \lesssim z \lesssim 2.5$~Mm), the absolute contribution increases despite the reduced filling factor. This implies that the rise in $\langle Q_{\rm mech}\rangle_{\rm sh}$ more than compensates for the smaller spatial coverage of shocks. This behaviour cannot be attributed solely to a selection effect. The increase in the absolute shock contribution demonstrates that shock-related dissipation becomes intrinsically stronger at larger $B_y$, even when integrated over the full horizontal extent. This points to stronger and/or more dissipative perturbations in the upper chromosphere, consistent with magnetic modification of wave propagation, channelling, and mode coupling in the low-$\beta$ regime we previously discussed, leading to fewer but more energetic magneto-acoustic shocks.

\subsubsection{Current sheets}

Despite the overall decrease in the CS filling factor with height as the plasma $\beta$ decreases below 1, we note two different regimes when it comes to comparing the different simulations and thus the impact of the injected $B_y$ in the bottom-left panel. As the latter increases, the filling factor first diminishes in most of the chromosphere ($z\lesssim 2$~Mm) before starting to increase higher up. The overall reduction of $\beta$ throughout the low-chromospheric layers of \textit{By200} (red) and \textit{By800} (green) strengthens the retro-action of the Lorentz-force on plasma flows, making the field less susceptible to twisting and therefore less prone to forming new small-scale CSs. Nevertheless, the trend reversal at higher altitude ($z\gtrsim 2$~Mm) acknowledges the particularity of \textit{By200} and \textit{By800} quasi-static regimes. In those, the accumulated horizontal field, loaded by flux emergence, forms a dynamic network of low-lying loops that have been randomly shuffled on the way, which will further promote non-parallel interaction with the newly emerging flux. These interactions produce extended current layers and long-lived reconnection sites, consistent with earlier studies of chromospheric reconnection and flux emergence-driven CSs \citep{archontisCLUSTERSSMALLERUPTIVE2014,hansteenBombsFlaresSurface2017,hansteenEllermanBombsUV2019,robinsonIncoherentFieldCoherent2022}.

When looking at the mean mechanical heating associated with reconnecting CS events $\langle Q_{\rm mech}\rangle_{\rm cs}$ in the bottom-middle panel, this increases consistently across the chromosphere as a function of the injected $B_y$. This is further pronounced as the $\beta\gtrsim 1$ regime is reached. Even though we have seen reconnection sites become less common below $z\sim2$~Mm, their local heating rate does indeed grow substantially in this regime. The decrease in $\beta\propto e_{\rm int}/e_{\rm mag}$ enhances the magnetic energy density, $e_{\rm mag}$, relative to the internal energy, $e_{\rm int}$, which favours strong transfer via ohmic dissipation. Flux emergence further drives reconnection by forcing interactions between the rising field and the pre-existing chromospheric network, which not only increases ohmic heating, but also compressive and viscous contributions generated by reconnection outflows (see also \citetalias{norazChromosphereQuietSun2026}). Together, these effects lead to a robust enhancement of reconnecting-CS-driven mechanical heating across the upper chromosphere, as further confirmed by the evolution of their absolute contribution to $Q_{\rm mech}$ in the bottom-right panel.

\subsection{Small-scale dynamics changes}\label{sec:horiz}

To illustrate how enhanced magnetic fields modify chromospheric small-scale dynamics, Fig.~\ref{fig:HorizLoc} shows temperature maps at three representative heights in the \textit{By800} model, namely $z = 0.5$, 1.0, and 1.5~Mm (left to right). Only one quarter of the horizontal domain is displayed to emphasise fine-scale structuring. Shocks (purple) and CSs (green), are overlaid on greyscale temperature maps, where brighter regions correspond to hotter plasma. We will discuss it in direct comparison with the \textit{Ref} case, for which we refer the interested reader to \citetalias{norazChromosphereQuietSun2026} for details.

At $z = 0.5$~Mm, the thermal morphology remains broadly similar to the reference case. CSs are already present but shocks are rare due to the relatively low average Mach number at this height (see also Fig.~\ref{fig:ShCsHeating}). CSs are preferentially spreading horizontally at this height due to the convective overturn and produce relatively thick overlays here, even though the underlying structures remain intrinsically thin \citepalias[e.g.][see also Eq.~\ref{eq:alpha_crit}]{norazChromosphereQuietSun2026}.

At $z = 1.0$~Mm, shocks become clearly visible, although their filling factor remains limited in comparison to \textit{Ref}, as can be expected from the top-left panel in Fig.~\ref{fig:ShCsHeating}. They correlate well with local temperature enhancements, confirming their role in intermittent small-scale heating already discussed in \citetalias{norazChromosphereQuietSun2026}. However, a key difference is that CSs now also coincide with hot (brighter) regions at this height, which was only observed higher in the atmosphere of \textit{Ref}. In \textit{By800}, the horizontally averaged plasma-$\beta$ is systematically reduced throughout the chromosphere, so that magnetic energy is comparable to internal energy ($\beta\sim 1$) already at $z\sim 1$~Mm. As a result, ohmic dissipation associated with CSs is no longer energetically constrained (i.e. $e_{\rm mag}\gtrsim e_{\rm int}$) and can visibly imprint the local temperature structure.

\begin{figure*}
    \begin{center}
        \includegraphics[width=0.33\linewidth]{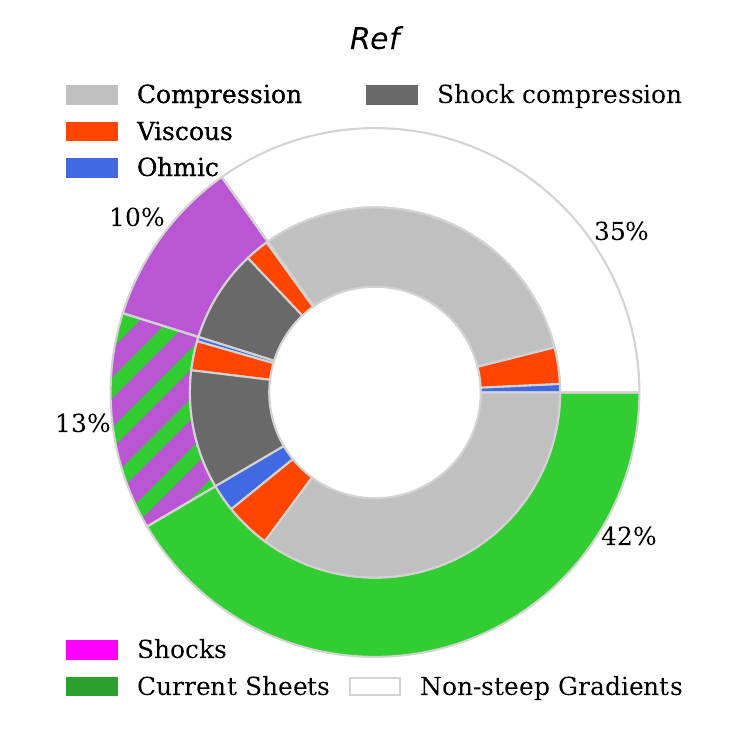}
        \includegraphics[width=0.33\linewidth]{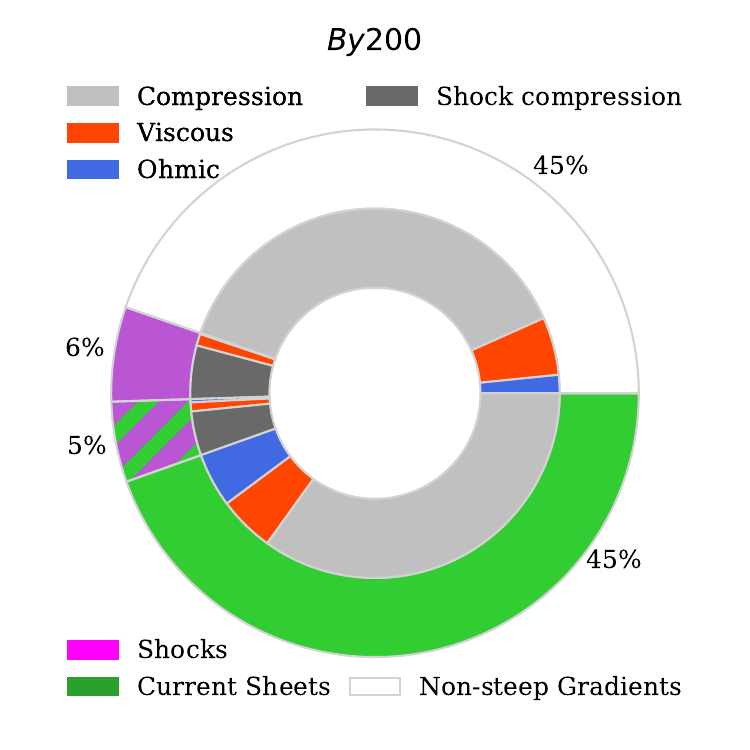}
        \includegraphics[width=0.33\linewidth]{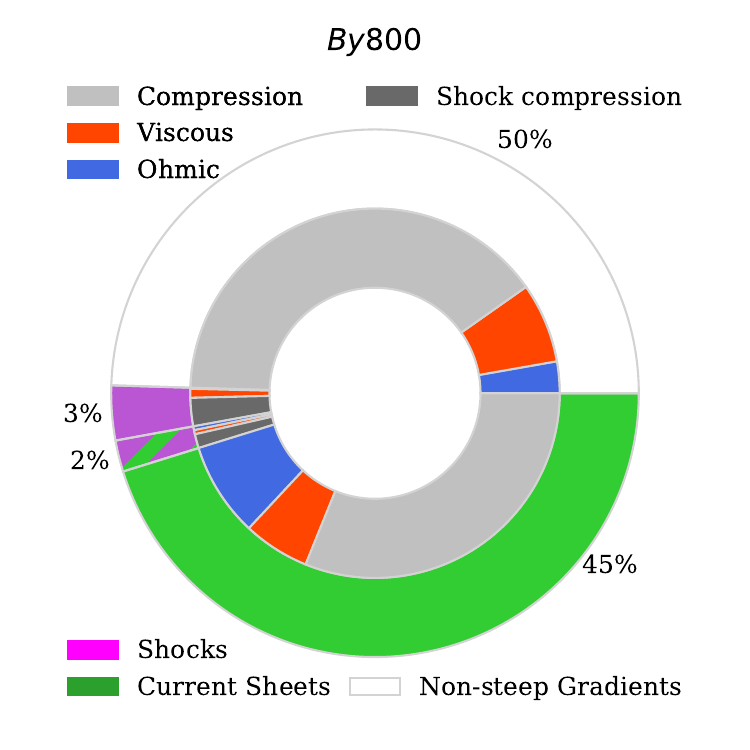}
    \end{center}
    \caption{Relative contributions of shocks (purple), CSs (green), and non-steep gradients (white) to the integrated mechanical heating of the chromosphere ($Q_{\rm mech}=Q_\nu + Q_\eta + Q_{comp}$, in red, blue, and grey, respectively). We present it for the three runs studied here: \textit{Ref} (\textit{left}), \textit{By200} (\textit{middle}), and \textit{By800} (\textit{right}). The different profiles used are spatially averaged over the chromospheric extent defined in the text body, with the outer ring indicating the physical processes involved (shocks, CSs, or neither), and the inner ring explicitly showing the associated dissipation mechanisms: viscous (red), ohmic (blue), and compression (grey). A darker shade of grey is used to highlight the shock compression contribution (see also Appendix A of \citetalias{norazChromosphereQuietSun2026}). The hatched segments indicate the contribution of regions where both shocks and CSs overlap.}\label{fig:PieCharts}
\end{figure*}

At $z = 1.5$~Mm, the dynamics become strongly magnetically organised. The temperature and dissipation patterns are elongated predominantly along the y-direction, reflecting the imposed strong $B_y=800$~G at the lower boundary of the domain. This large-scale magnetic configuration now visibly channels shock propagation (see animation attached to Fig.~\ref{fig:HorizLoc}) and reshapes the flow morphology through Lorentz-force feedback. The average plasma-$\beta$ at this height is about an order of magnitude smaller than in \textit{Ref}, reaching about 0.1 on average, which explains the dominant magnetic control of the chromospheric structuring.

\subsection{Summary of the contributions}

To give an overall summary of the different contributions to the mechanical heating over the whole chromosphere, we define the chromospheric boundaries following the approach proposed in \citetalias{norazChromosphereQuietSun2026}. The bottom height is set by reporting where the photospheric radiative equilibrium approximation breaks \citep{schwarzschildEquilibriumSunsAtmosphere1906}, which occurs, in practice, at $z=600$, 570, and 530~km in \textit{Ref}, \textit{By200}, and \textit{By800}, respectively. The top of the chromosphere is subsequently approximated where the horizontally averaged temperature reaches 20~kK, following the common proxy for the substantial decrease in H$\alpha$ emissions. This is met at $z = 2.57$, 2.35, and 2.11~Mm. We further discuss this later change and its implications in Sect.~\ref{sec:sect4}.

As already presented for \textit{Ref} in \citetalias{norazChromosphereQuietSun2026}, we integrated the relevant mechanical heating terms $Q_{\rm mech}=Q_\nu+Q_\eta+Q_{\rm comp}$ over the defined chromospheric extent and show them in Fig.~\ref{fig:PieCharts} for the different runs. The resulting energy budget is summarised as a pie chart, with the outer ring indicating the heating processes (shocks, CSs, or neither) and the inner ring specifying the deposition term contributions.

The relative contribution of shocks (purple) to the mechanical heating decreases with increasing injected magnetic field strength, dropping from about one fourth in the weakly magnetised QS \textit{Ref} case to about 5\% in the strong \textit{By800} model. In contrast, heating associated with reconnecting CSs (green) consistently accounts for about half of the total budget (55 to 47\%). Combined with the increase in total mechanical heating reported in Fig. \ref{fig:Qmech}, these results indicate that, although the filling factor of CSs decreases, their local heating efficiency increases substantially, as expected from the bottom-right panel of Fig.~\ref{fig:ShCsHeating}. This enhancement is primarily driven by the larger magnetic energy reservoir and its subsequent release through slow diffusion and reconnection-dynamics deposition within CSs, as illustrated further in Sect. \ref{sec:mass_loading}. The substantial increase in the relative Ohmic heating (blue) in CSs contributions is also consistent with observational trends when going towards more-active regions \citep{morosinSpatiotemporalAnalysisChromospheric2022}.

We can also note that the contribution of non-steep gradient (white part) increases as well from \textit{Ref} to \textit{By800}. This contribution likely comes from the energy deposition of broader current layers, but also the propagation of high-amplitude and linear waves (see the animation attached to Fig.~3 in \citetalias{norazChromosphereQuietSun2026}). This is consistent with the scenario of an increased ramp effect, happening when the inclination of the field is more pronounced in the low solar atmosphere \citep{stangaliniMHDWaveTransmission2011}, subsequently allowing for a broader spectrum of waves to propagate in the upper atmosphere. A detailed characterisation of this transmission, and its quantitative impact on chromospheric and coronal coupling, lies beyond the scope of the present study (see e.g. \citealt{stangaliniDynamicsSmallscaleMagnetic2025,udnaesCharacteristicsAcousticwaveHeating2025,enerhaugIdentifyingMagnetohydrodynamicWave2025,cherryDecomposingWaveActivity2025}).

\section{Atmospheric coupling}\label{sec:sect4}

\subsection{Increased density scale height}\label{sec:Hrho}

\begin{figure}
    \begin{center}
        \includegraphics[width=\linewidth]{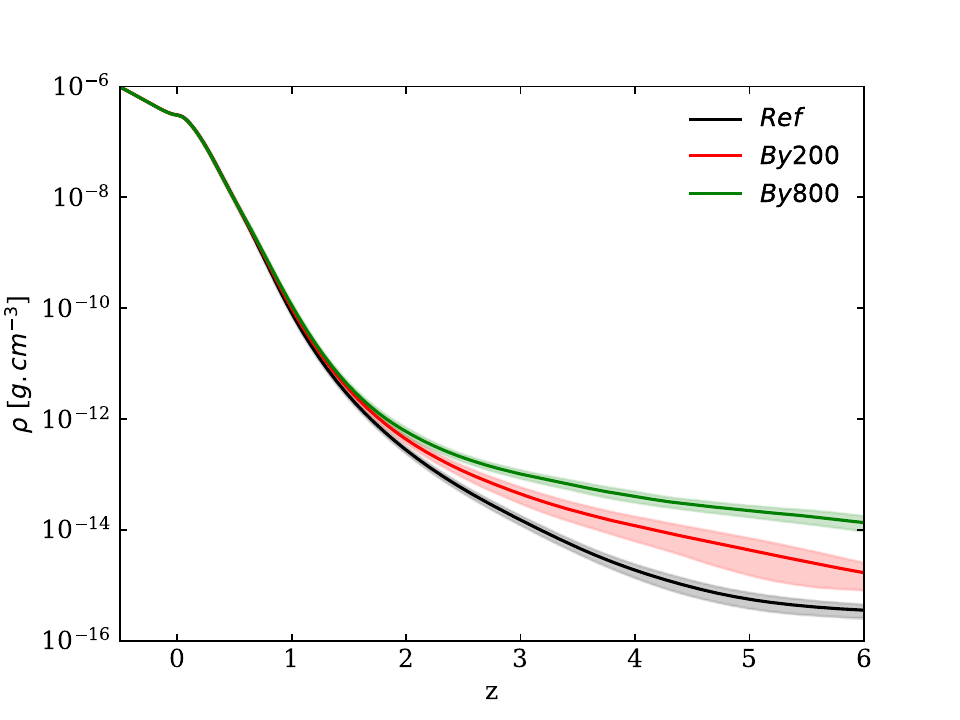}
    \end{center}
    \caption{Comparison of density profiles among \textit{Ref}, \textit{By200} and \textit{By800} in black, red, and green, respectively, averaged horizontally in space and over one solar hour in time. The envelope indicates $\pm 1$ standard deviation in time.}\label{fig:atmoResp}
\end{figure}

\begin{figure*}
    \begin{center}
        \includegraphics[width=0.5\linewidth]{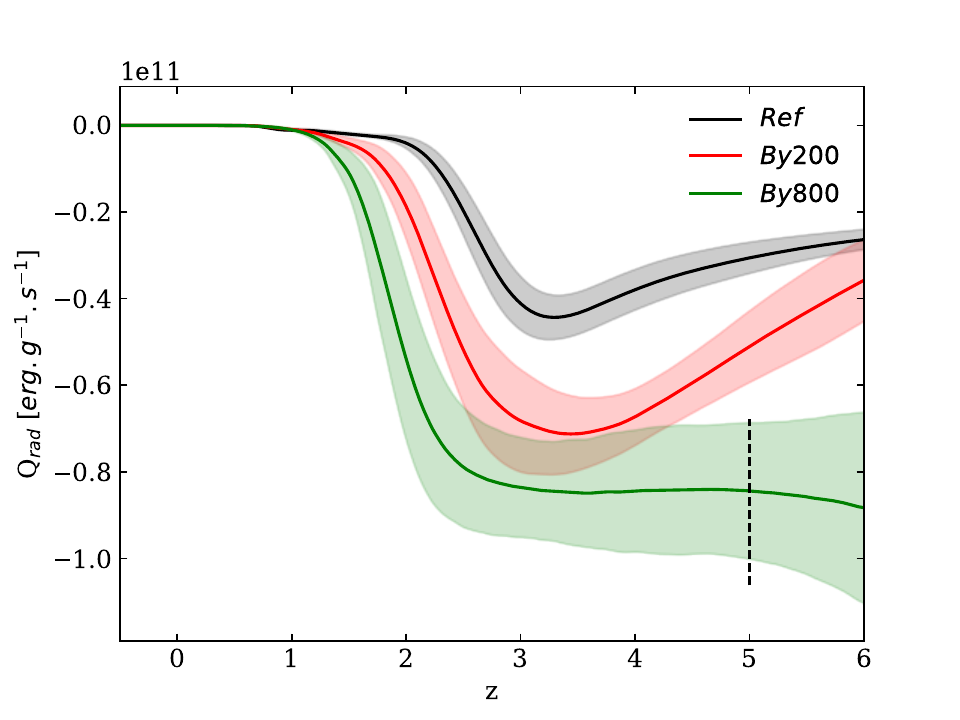}
        \includegraphics[width=0.48\linewidth]{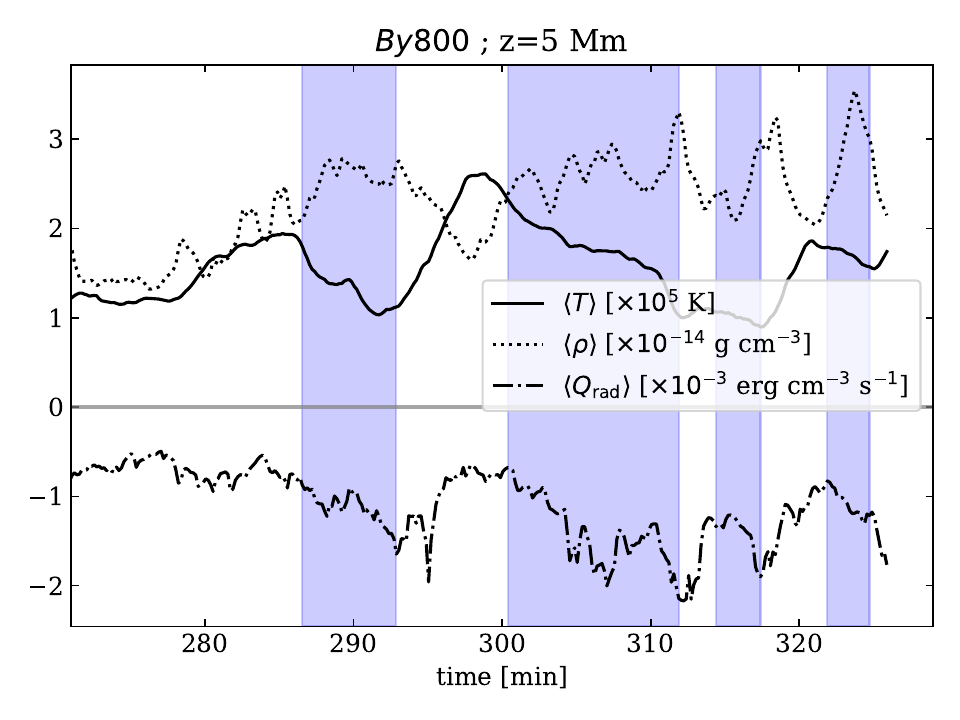}
    \end{center}
    \caption{\textit{Left:} Comparison of radiative cooling profiles. The layout is similar to Fig.~\ref{fig:atmoResp}. \textit{Right:} Temporal evolution of temperature $T$ (solid), density $rho$ (dotted) and radiative cooling $Q$ (dotted-dashed) values, averaged over the horizontal extent at $z=5$~Mm. Blue-shaded time ranges highlight periods when the temperature decreases substantially, in order to compare with radiative cooling and density enhancements.}\label{fig:tempCorr}
\end{figure*}

Understanding the non-monotonic behaviour of $T_{\rm cor,bot}(B_z)$, despite the monotonic increase in $T_{\rm chromo}(B_z)$, requires examining the coupling between the chromosphere and corona. To this end, and to elucidate the coronal temperature decrease in the \textit{By800} run observed in Fig. 4, we show horizontally and temporally averaged profiles of density and radiative cooling in Figs.~\ref{fig:atmoResp} and ~\ref{fig:tempCorr}, respectively. Both quantities increase consistently in amplitude at all heights as a function of the magnetic field amplitude injected (from black to green).

In Fig.~\ref{fig:atmoResp}, the density profiles of \textit{By200} and \textit{By800} start to deviate from the \textit{Ref} profile already in the upper chromosphere, around $z\sim 2$~Mm, leading monotonically to denser plasma higher up in comparison to \textit{Ref}. This deviation is accompanied by a change in the slope of the profiles, indicating a sudden increase in the density scale height, $H_\rho=\Delta z/\Delta \ln{\rho}$. In the lower chromosphere ($0.5<z<1.5$~Mm), $H_\rho\sim$ is consistent with $0.12$~Mm for the 3 models, whereas the profiles become nearly flat above $z\sim 5$~Mm, implying a scale height exceeding the vertical extent left up to the top of the simulated domain. We can understand this behaviour as $H_\rho\propto T$ under hydrostatic approximation. Although the hydrostatic approximation may appear restrictive given the highly dynamic, small-scale nature of the simulations, it appears to remain consistent once quantities are averaged over space and time in the context of this work.

As a result, the onset of the chromospheric temperature rise leads, to first order, to an increase in $H_\rho$ and sets the thermodynamic conditions at the coronal base by increasing density. This highlights the importance of chromospheric structure, topology, and thermodynamics in energy and mass transport up to the lower solar corona.

\subsection{Enhanced radiative cooling}

In the left panel of Fig.~\ref{fig:tempCorr}, we show a comparison of the radiative cooling profiles $Q_{\rm rad}$. We include radiative losses occurring from the top of the chromosphere upwards; namely, those computed using the semi-empirical radiative-loss recipes of \cite{carlssonApproximationsRadiativeCooling2012} for hydrogen, calcium, and magnesium, together with optically thin radiative losses based on CHIANTI atomic data \citep{dereCHIANTIAtomicDatabase1997,landiCHIANTIAtomicDatabase2006}. Here, we express $Q_{\rm rad}$ per unit mass, thereby limiting the dominance of high-density regions seen for \textit{By800}. Figure ~\ref{fig:atmoResp} offers a meaningful comparison, along with a consideration of the possibility of erg/s/cm3 based on the plot.

Above $z\sim 1.5$~Mm, the absolute amplitude of $Q_{\rm rad}$ increases with the imposed magnetic-field strength. This behaviour is consistent with the enhanced density stratification discussed in the previous section, since under optically thin and fully ionised conditions, appropriate for the corona, radiative losses scale as $Q_{\rm rad}\propto \rho^2$ \citep[e.g.][]{mihalasFoundationsRadiationHydrodynamics1984,2003rtsa.book.....R}. The pronounced enhancement of radiative cooling in the \textit{By800} run relative to \textit{Ref} is particularly noteworthy, given that presenting the losses per unit mass already mitigates the impact of the highest density regions.

To further demonstrate that the density increase, and the resulting enhancement of radiative cooling, is the primary driver of the temperature decrease observed at the coronal base in Fig.~\ref{fig:tempProfs}, we examine the temporal evolution of temperature, density, and radiative cooling in the right panel of Fig.~\ref{fig:tempCorr}, after a horizontal averaging at $z=5$~Mm in the \textit{By800} simulation. Blue-shaded intervals indicate periods of pronounced temperature decrease. These episodes coincide with enhanced radiative cooling rates (dash-dotted curve) and periods of high or an increase in densities (dotted curve).

\begin{figure*}
    \begin{center}
        \includegraphics[width=\linewidth]{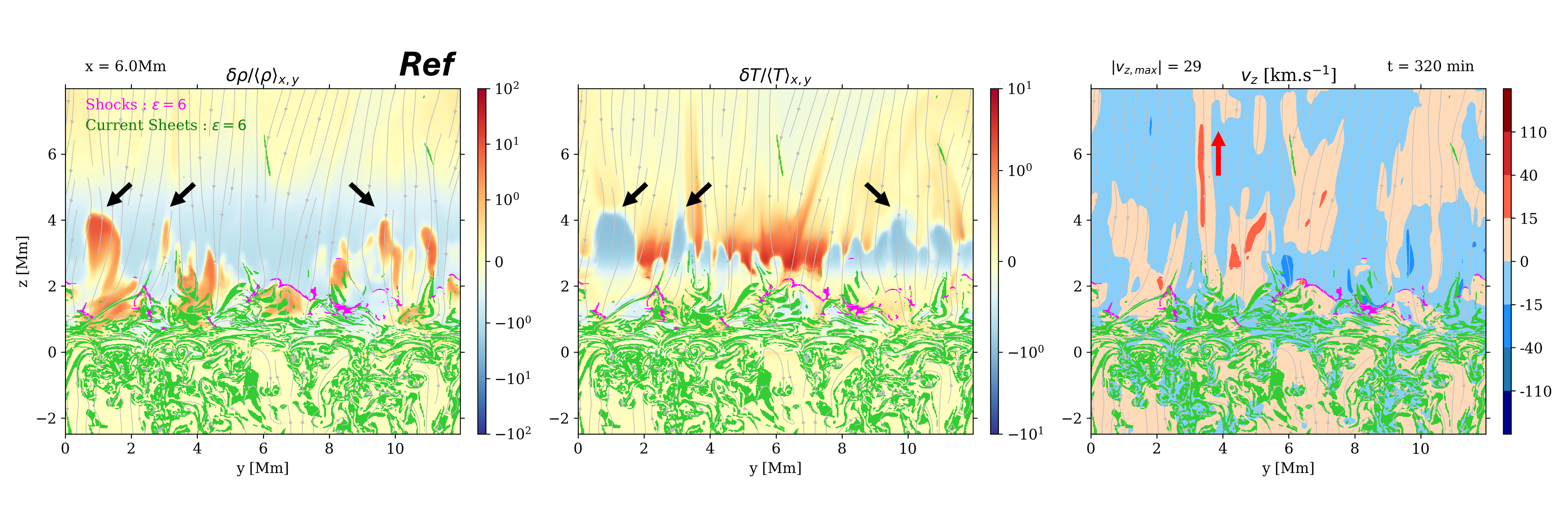}
        \includegraphics[width=\linewidth]{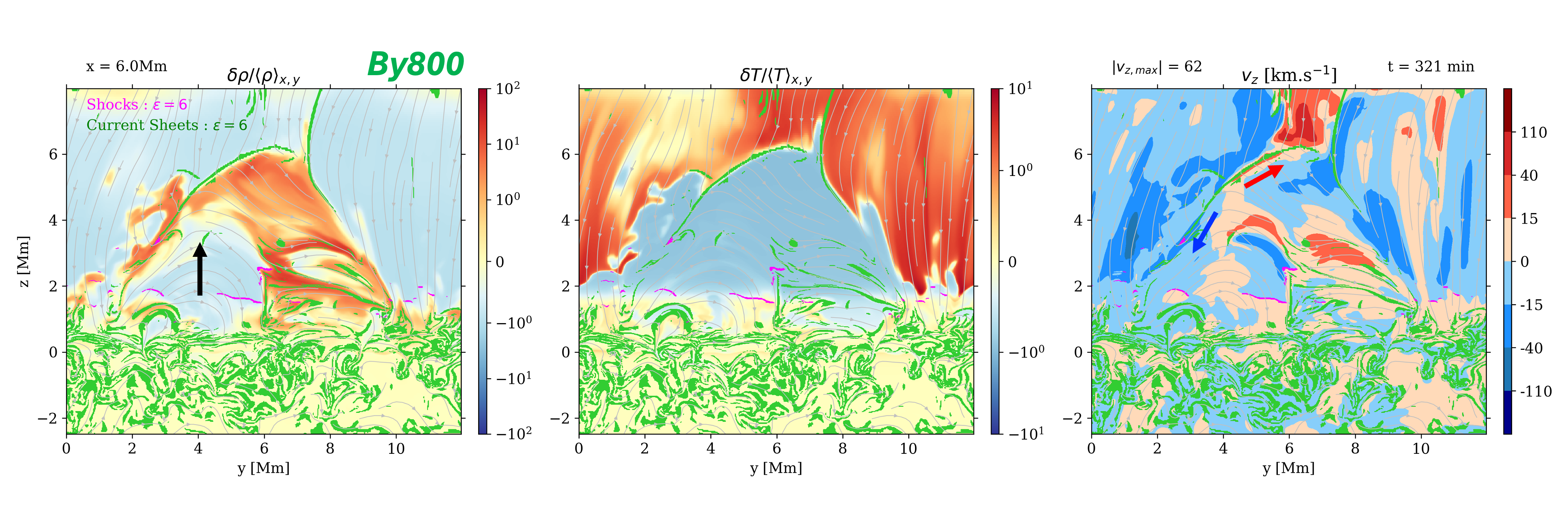}
    \end{center}
    \caption{Mass-loading behavior of \textit{Ref} (top row) and \textit{By800} (bottom row). We highlight the position of shocks and CSs for all panels, following Eqs.~\ref{eq:cs_crit} and \ref{eq:alpha_crit}, respectively. \textit{Left:} Density variation, $\delta\rho/\langle\rho\rangle_{x,y}$, taken at $x=6$~Mm for each given time step and of each given simulation. This illustrates material more (red) or less (blue) dense than the surrounding material at that height. \textit{Middle:} Same but for the temperature variation, $\delta T/\langle T\rangle_{x,y}$. \textit{Right:} Same but for the vertical velocity component, $v_z$, where red (blue, respectively) shows upwards (downwards, respectively) motions. Please note for the top row that dark arrows indicate over-densities, corresponding here to cooler material entering the coronal medium, and corresponding to shock-mediated (purple contours) type-I spicule dynamics (see also the red arrow indicating an upwards spicular motion and also see Fig.~4 of \citetalias{norazChromosphereQuietSun2026}). We also note that for the bottom row, the dark arrow highlights the motion of the emerging magnetic loop-like structure, transporting the over-density (red) up to coronal heights. Interactions with the overlying magnetic field create thin reconnecting CS structure (green contours), where bipolar flows are highlighted by red and blue arrows (see also Fig.~5 of \citetalias{norazChromosphereQuietSun2026}). Associated movies are available online.}\label{fig:mass_loading}
\end{figure*}

The correlations shown here are particularly relevant given that all quantities are averaged over the full horizontal extent of the domain. They reveal a clear causal imprint of density enhancements on radiative cooling, as expected, and well as, in turn, on the global temperature evolution at this height. It is important to recall that other non-local transport processes, such as thermal conduction and advection, can also contribute to coronal cooling at coronal heights. However, these processes fluctuate rapidly between heating and cooling and exhibit no clear temporal correlation with the mean temperature decreases, even when examined across multiple coronal heights. 

Another aspect that can contribute to the non-monotonic coronal temperature response is the dominant orientation of the magnetic field, which varies from one model to another (see Fig.~\ref{fig:3Drend}). A loss of magnetic connectivity to the lower atmosphere would modify the redistribution of heat. However, such a configuration should instead lead to higher coronal temperatures, as the deposited energy could no longer be efficiently transported downwards by thermal conduction, which was not observed for \textit{By800}. Enhanced density-driven radiative losses therefore emerge as the dominant cooling mechanism governing the global temperature-decrease episodes and the reduced mean coronal-base temperature of \textit{By800} relative to \textit{Ref}, as seen in Fig.~\ref{fig:tempProfs}.

\subsection{Mass-loading}\label{sec:mass_loading}

We have shown that the overall temperature decrease at the base of the corona in \textit{By800} relative to \textit{Ref} is primarily driven and sustained by an increase in density at that height. This naturally raises the question of how mass is effectively transported from the chromosphere into the low corona. Addressing this question quantitatively is beyond the scope of the present paper; nonetheless, our aim here is to propose a qualitative analysis of the simulated dynamics to illustrate the mechanisms of low-atmospheric coupling and further guide the quantification of mass fluxes in future works.

In Fig.~\ref{fig:mass_loading}, we illustrate several diagnostic quantities that characterise and compare the plasma dynamics in the \textit{Ref} (top row) and \textit{By800} (bottom row) cases. In the \textit{Ref} simulation, we observe recurrent type-I spicule dynamics, characterised by upwards motions (see red arrow in the right panel) of dense (red patches in the left panel) and cool (blue patches in the middle panel) chromospheric plasma. The animation shows that these spicules are shock-driven and guided by the predominantly vertical magnetic field, in agreement with previous analyses of the same run \citep[Fig.~4 of][]{norazChromosphereQuietSun2026} and with observational interpretations \citep{hansteenDynamicFibrilsAre2006,depontieuHighResolutionObservationsModeling2007}. We see in the left panel that this spicular dynamics largely contribute to chromosphere-to-corona mass-loading, together with magnetic swirling motions, as also illustrated in Fig.~1 of \cite{norazChromosphereQuietSun2026}. Although a quantitative assessment of their respective contributions to mass and energy transport lies beyond the scope of this study, these processes are expected to play a key role in quiet-Sun atmospheric coupling \citep[see e.g.][]{martinez-sykoraGenerationSolarSpicules2017,finleyStirringBaseSolar2022,breuSwirlsSolarCorona2023,skirvinPoyntingFluxMHD2024,chandraStatisticalPropertiesSpicules2025}. In particular, the role of spicules in this coupling remains under debate \citep{klimchukKeyAspectsCoronal2015,sowmondalContributionSpiculesSolar2022}.

In contrast, the bottom row reveals a notably different magnetic and dynamical regime in \textit{By800}. We recall here that $\delta\rho(y,z)/\langle\rho\rangle(z)$ is the relative difference with respect to the horizontally averaged density at this height, $z$, and that this reference value has increased in \textit{By800} (see Fig.~\ref{fig:atmoResp}). The magnetic field now exhibits a strong horizontal component resulting from the imposed flux injection, as discussed in Sect.~\ref{sec:sect2}. At granular scales, this manifests as the emergence of low-lying magnetic loops (black arrow in the bottom-left panel). As these loops rise through the chromosphere, their upper segments trap and advect cool, dense chromospheric plasma upwards into the corona, as evidenced by the co-spatial red and blue patches in the bottom-left and bottom-middle panels, respectively (see the attached animation). This process contributes efficiently to increasing the density at the base of the corona, consistent with similar mechanisms identified in more magnetically active simulations \citep{druettFormationHeatingChromospheric2022}.

We stress that the increase in horizontally averaged density in Fig.~\ref{fig:atmoResp} reflects a combination of enhanced heating and direct mass injection associated with flux emergence, both contributing to the coronal mass-loading. To further assess this point, we analysed regions of \textit{By800} without flux emergence and still found enhanced heating and a systematic increase in column mass with respect to the reference case. This indicates that the density increase is not solely driven by the emergence-related structure, but that it also reflects a more broadly global thermodynamic response.

Finally, we can note from the animation that the top of the dome-like magnetic structure interacts continuously with the overlying magnetic field during its emergence. In the bottom-left panel, we indicate upwards and downwards plasma motion associated with the reconnecting structure, with a red and blue arrow, respectively. The corresponding thin green overlay between the ambient field and magnetic dome structure can track the core part of the CS that is dynamically relevant for reconnection (thanks to Eq.~\ref{eq:cs_crit}) where these bipolar flows originate. The upwards jet gives birth to a high-speed surge, as acknowledged by the red patch at the top of the right panel, between $y=6$ and 8~Mm. This type of chain of events is recurrent in \textit{By800} and should be further characterised in future comparisons with observational constraints \citep[see e.g.][]{heyvaertsEmergingFluxModel1977,yokoyamaMagneticReconnectionOrigin1995,isobeFilamentaryStructureSun2005,isobeConvectiondrivenEmergenceSmallScale2008,nobrega-siverioSmallscaleMagneticFlux2024,huangHighresolutionObservationsSmallscale2026}.

\section{Discussion}\label{sec:sect5}

As illustrated by the animations associated with Figs.~\ref{fig:3Drend} and \ref{fig:mass_loading}, small-scale flux-emergence events that transport chromospheric plasma to higher atmospheric layers appear to be a key mechanism for explaining the enhanced density at the base of the corona in \textit{By800}. However, it should be noted that the global increase in temperature and the resulting modification of the pressure gradient across the horizontal extent of the domain might also contribute to driving this mass-loading. However, quantifying the relative contribution of these processes is beyond the scope of the present study and would require a dedicated analysis similar to that proposed by \cite{druettFormationHeatingChromospheric2022}.

The flux-emergence scenarios considered here are deliberately idealised, relying on the injection of untwisted horizontal magnetic fields at the lower boundary. While this approach is well-suited for controlled parametric exploration, future studies should aim to incorporate more realistic boundary conditions, either driven by observations \citep[e.g.][]{chenDatadrivenRadiativeMagnetohydrodynamics2025a} or self-consistently coupled to global convection and dynamo models \citep[e.g.][]{fangDYNAMICCOUPLINGCONVECTIVE2012}.

Because the magnetic flux is continuously injected, the system does not relax towards a passive post-emergence state but instead approaches a quasi-stationary, driven regime, in which heating, mass-loading, and radiative cooling reach a dynamic balance. Although a slow secular evolution remains visible (see Fig.~\ref{fig:atmoCA}), the main thermodynamic trends discussed here are established after the initial transient phase and remain robust, even at later stages of \textit{By200} simulated duration.

We want to stress here that the coronal temperatures obtained in this set of simulations should not be interpreted as definitive predictions. As demonstrated here, the coronal thermal structure strongly depends on how chromospheric heating and energy dissipation are modelled. Previous studies have shown that additional physical ingredients, such as ion–neutral interactions and non-equilibrium ionisation, can significantly modify chromospheric heating efficiencies and loop thermodynamics \citep{martinez-sykoraTWODIMENSIONALRADIATIVEMAGNETOHYDRODYNAMIC2012,shelyagHEATINGPARTIALLYIONIZED2016,nobrega-siverioAmbipolarDiffusionBifrost2020,martinez-sykoraIonNeutralInteractions2020}. These effects are available within Bifrost and should be explored in future extensions of the present work. Furthermore, coronal temperatures are also expected to depend on magnetic topology and spatial scale. Larger scale or more active configurations have been shown to produce hotter, million-degree coronae (\citealt[][]{carlssonPubliclyAvailableSimulation2016,finleyStirringBaseSolar2022}, see also the conclusion of \citealt{przybylskiStructureDynamicsInternetwork2025}), underscoring the need to extend parametric studies to a broader range of magnetic environments.

We conservatively restricted our analysis to heights below 7~Mm. Above $\sim 7.5$~Mm, the solution becomes increasingly sensitive to the top boundary condition. However, all trends discussed in this study, including the non-monotonic coronal response, are established well below this region and are not affected, either qualitatively or quantitatively, by measurable boundary effects in the analysis performed.

\section{Conclusion and perspective}\label{sec:sect6}

In this work, we conducted a parametric 3D radiative-MHD study of quiet-Sun atmospheric coupling with Bifrost, building on the reference simulation presented in \citetalias{norazChromosphereQuietSun2026}. By injecting horizontal magnetic flux of increasing amplitude into the sub-surface convection zone, we constructed two additional models, spanning weakly magnetised, coronal-hole-like conditions (labelled \textit{Ref} in the paper) to more typical QS amplitudes with intermediate and strong small-scale flux emergence (referred to as \textit{By200} and \textit{By800}; Fig.~\ref{fig:3Drend}). All the simulations reached a quasi-static state into which the magnetic flux both enters and leaves the computational domain, enabling a direct comparison of their thermodynamic and dynamical properties.

The chromosphere and corona respond differently to increasing magnetic-field amplitude. While the chromospheric temperature increases monotonically with the amplitude of the emerging flux imposed, the temperature at the base of the corona shows a non-monotonic response, first rising in the intermediate case relative to the reference case and then decreasing in the strongly magnetised case (see Fig.~\ref{fig:tempProfs}). This contrast motivated an analysis of chromospheric heating and atmospheric coupling.

In Sect.~\ref{sec:sect3} we show that the total mechanical chromosheric heating increases with magnetic-field strength, driven primarily by reconnecting current sheets, which consistently contribute about half of the heating (55–47\%). Although their filling factor decreases in most of the chromosphere, their local heating efficiency increases as additional magnetic energy is injected, due to the associated reduction in plasma-$\beta$. In contrast, shock-driven heating becomes progressively less important, from 23\% in \textit{Ref} to 5\% in \textit{By800}. Taken together, stronger magnetic fields promote more efficient chromospheric heating in our QS models.

Despite this enhanced heating, Sect.~\ref{sec:sect4} shows that the coronal-base temperature decreases in the strongly magnetised case, due to a substantial density increase. Enhanced chromospheric heating increases the density scale height, thereby setting a higher density at the base of the corona, through the combined effect of both heating-induced and direct mass-loading from flux emergence. This density increase strongly amplifies radiative losses, which dominate the cooling in coronal energy balance and lead to global temperature-decrease episodes. The cooler coronal-base temperatures, observed in the strongly magnetised case, therefore do not reflect reduced heating efficiency, but a density-controlled equilibrium tightly linked to the redistribution of energy and mass across atmospheric layers.

These results highlight the central role of chromospheric temperature in setting the density scale height, which in turn constrains the density supplied to the corona and subsequent thermal equilibrium. Low atmospheric heating and mass-loading thus emerge as key regulators of coronal thermodynamics, even under increased magnetic activity. This has direct implications for surface-to-corona coupling in solar-wind models. While frameworks such as Wang–Sheeley–Arge relate surface magnetic fields to wind properties \citep{wangLatitudinalDistributionSolarwind1990,argeImprovementPredictionSolar2000,argeStreamStructureCoronal2004}, they still rely on simplified low-atmosphere parameterisations, despite a strong sensitivity to lower-boundary conditions \citep[e.g.][]{kuzmaCOCONUTNovelFastconverging2023}. Our results indicate the need for an explicit incorporation of chromospheric heating and mass-loading in these parameterisations.

\begin{figure}
    \begin{center}
        \includegraphics[width=\linewidth]{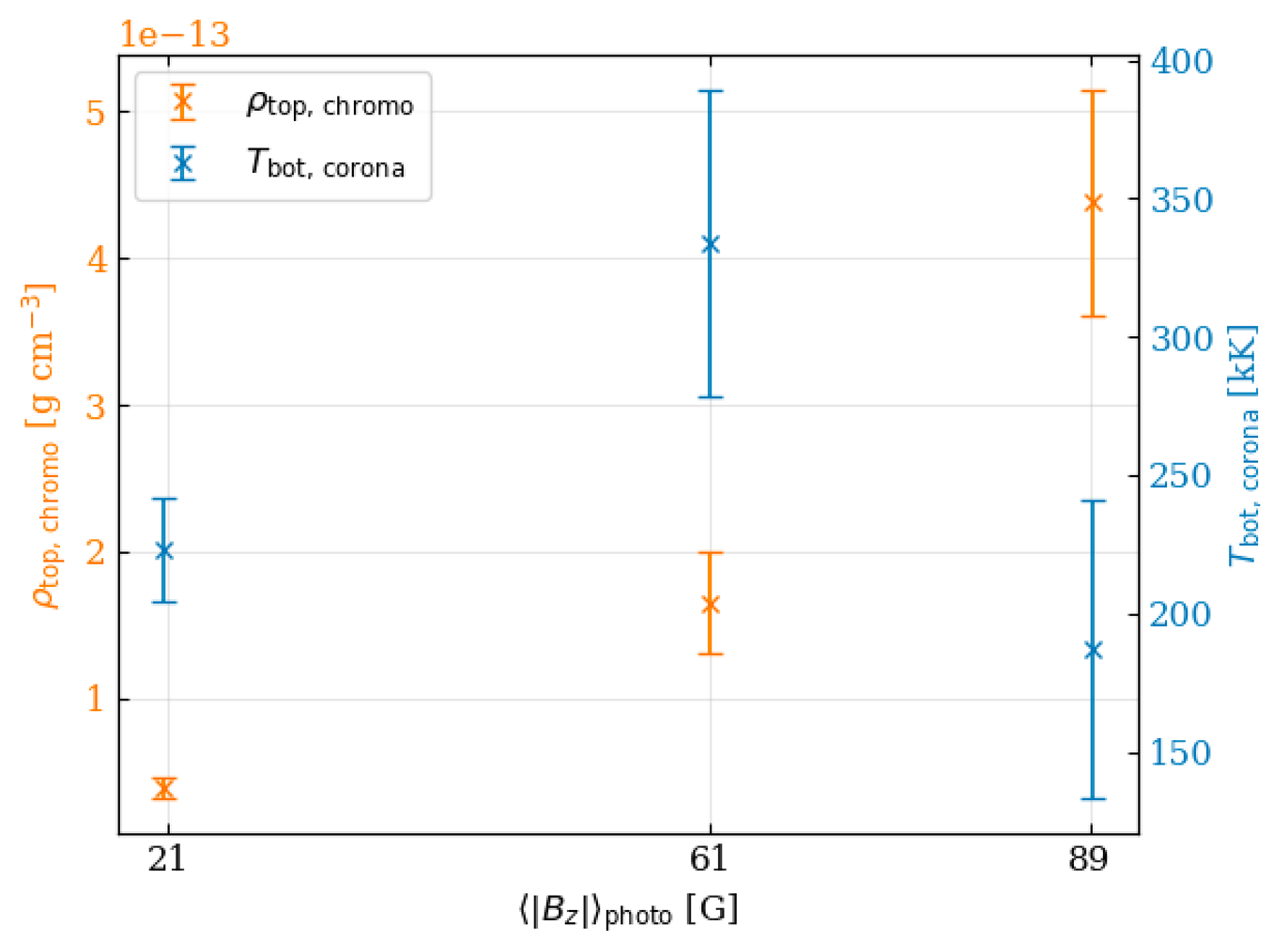}
    \end{center}
    \caption{Density at the top of the chromosphere, $\rho_{\rm top, chromo}$ (orange symbols, left axis), and temperature at the base of the corona, $T_{\rm bot, corona}$ (blue symbols, right axis), as functions of the mean unsigned photospheric vertical magnetic field,$\langle |B_z|\rangle_{\rm photo}$. The chromospheric density is averaged at the height where the horizontally averaged temperature reaches 20~kK, while the coronal temperature is averaged at $z=6$~Mm. Error bars indicate one standard deviation in time over the selected quasi-static interval. The figure highlights the monotonic increase in chromospheric density with magnetic-field strength and the non-monotonic response of the coronal-base temperature.}\label{fig:takeHome}
\end{figure}

Figure~\ref{fig:takeHome} summarises the density at the top of the chromosphere and the temperature at the coronal base as functions of the mean unsigned photospheric field, $\langle |B_z|\rangle_{\rm photo}$. While $\rho_{\rm top, chromo}$ increases monotonically with $\langle |B_z|\rangle_{\rm photo}$, $T_{\rm bot, corona}$ exhibits a non-monotonic response, with a maximum at intermediate field strength followed by a decline beyond a threshold value. We stress that the three simulations considered here do not constitute a dense parametric survey, but rather a controlled parametric exploration aimed at isolating the underlying physical mechanisms. Assessing the robustness and physical origin of this trend over a broader parameter space is a key objective of future work.

The strong density-driven radiative cooling reported here also implies enhanced emission signatures potentially observable with the current instrumentation \citep[e.g.][]{robinsonTracingSignaturesQuiet2023}. Future comparisons with AIA and Solar Orbiter observations, as well as future facilities such as EST \citep{quinteronodaEuropeanSolarTelescope2022} and AtLAST \citep{wedemeyerMillimeterWavelengthObservationsActive2025} will help constrain chromospheric heating mechanisms and the mass–energy coupling between the lower solar atmosphere and the corona.

\begin{acknowledgements}
All authors are thankful to F. Zang, N. Poirier, B. Gudiksen, V. Hansteen, J. Mart\'inez-Sykora, K. Krikova and L. Rouppe van der Voort for useful discussions. The authors also thank the anonymous referee for useful and constructive remarks. We acknowledge funding support by the European Research Council (ERC) under the European Union’s Horizon 2020 research and innovation programme (grant agreement No 810218 WHOLESUN and No 101141362 Open SESAME), by the Research Council of Norway through its Centres of Excellence scheme (RoCS project number 262622), and through computational resources provided by Sigma2, the National Infrastructure for High Performance Computing and Data Storage in Norway. The work of GA was supported by the Action Thématique Soleil-Terre (ATST) of CNRS/INSU PN Astro, also funded by CNES, CEA, and ONERA. Data manipulation was performed using the numpy \citep{harrisArrayProgrammingNumPy2020} and the in-house \textit{Bifrost} analysis pipeline \textit{helita} python packages. Figures in this work were produced using the python packages matplotlib \citep{HunterMatplotlib} and pyvista \citep{sullivan2019pyvista}.
\end{acknowledgements}

\bibliographystyle{aa}
\bibliography{MyLibrary}

\end{document}